\documentclass[singlecolumn,amsmath,amssymb,nofootinbib,showkeys]{revtex4-1}
\usepackage[TS1,T1]{fontenc}
\usepackage{graphicx}
\usepackage{dcolumn}
\usepackage{bm}

\usepackage[mathlines]{lineno}

\usepackage{subcaption}
\usepackage{svg}
\usepackage{booktabs}
\usepackage{easyReview}
\usepackage{siunitx}
\usepackage{float}
\usepackage{amsmath,bm}
\usepackage{widetable}

\usepackage{todonotes}
\usepackage{menukeys}

\usepackage{nicefrac}

\newcommand{\lambdabi}{\Lambda_\textrm{shuttle}}
\newcommand{\nubi}{\nu_\textrm{shuttle}}
\newcommand{\Dbi}{D_\textrm{shuttle}}
\newcommand{\etabi}{\eta}

\newcommand{\shuttlespeed}{v_0}
\newcommand{\trainspeed}{v_\textrm{train}}

\newcommand{\emissions}{\mathcal{E}}
\newcommand{\quality}{\mathcal{Q}}
\newcommand{\trainEnergy}{e_\textrm{train}}
\newcommand{\shuttleEnergy}{e_\textrm{shuttle}}
\newcommand{\carEnergy}{e_\textrm{MIV}}

\newcommand{\fbi}{F}
\newcommand{\traffic}{\tilde \Gamma}
\newcommand{\trainocc}{\bm{\alpha}}
\newcommand{\funi}{(1-F)}
\newcommand{\meshsize}{\tilde{\ell}}
\newcommand{\dcut}{\bm{d_\textrm{c}}}
\newcommand{\dgdc}{\langle d \rangle_{d > \dcut}}
\newcommand{\dldc}{\langle d \rangle_{d < \dcut}}
\newcommand{\dgdctilde}{\langle \tilde{d} \rangle_{\tilde d > \tilde \dcut}}
\newcommand{\dldctilde}{\langle \tilde{d} \rangle_{\tilde d <  \tilde \dcut}}

\newcommand{\SuppInf}{Supporting Information}

\modulolinenumbers[5]

\begin{document}

\title{Sustainable and convenient: bi-modal public transit systems\\outperforming the private car}

\author{Puneet Sharma}
 \email{puneet.sharma@ds.mpg.de}
\author{Knut M. Heidemann}
 \email{knut.heidemann@ds.mpg.de}
\author{Helge Heuer}
\author{Steffen M\"uhle}
\author{Stephan Herminghaus}
\affiliation{Max-Planck Institute for Dynamics and Self-Organization (MPIDS), Am Fa\ss berg 17, 37077 G\"ottingen, Germany}

\date{\today}

\begin{abstract}
Mobility is an indispensable part of modern human societies, but the dominance of motorized individual traffic (MIV, i.e., the private car) leads to a prohibitive waste of energy as well as other resources. Here we show that by combining a line service (e.g., railway) system with a fleet of ride-pooling shuttles connecting line stops to desired pick-up and drop-off points, a bi-modal public transport system may result which provides on-demand door-to-door service at a service level (in terms of transit times) superior to current public transport, and with an overall comfort level comparable to MIV.  We identify the conflicting objectives for optimization, i.e., user convenience and energy consumption, and evaluate the system performance in terms of Pareto fronts. By means of simulation and analytical theory, we find that energy consumption can be as low as 20\ of MIV, at line service densities typically found in real settings. Surprisingly, we find favorable performance not only in urban, but also in rural settings.
\end{abstract}

\maketitle

\section{Introduction}

Transportation accounts for about one-fifth of global anthropogenic carbon emissions \cite{air_emission_review,air_emissions_Kontovas2016}, owing mainly to the fact that humans rely mostly on motorized individual vehicles (MIV) \cite{reliance_MIV_Douglas,reliance_MIV_newman1989cities}, i.e., private cars. Aside from the ensuing logistic \cite{parking_giuliano,manville2005parking,logistic_dongjoo,logistic_Jang,logistic_scott} and environmental \cite{eeaoccupancy,environment_joireman2004cares} problems of traffic congestion \cite{traffic_singapore,traffic,traffic_Poland,traffic_greenhouse} and air pollution \cite{eeaair,ashok_pollution}, MIV represents an enormous waste of energy. On average it amounts to moving about a ton of material \cite{mackenzie2014determinants} in order to move just one person \cite{tachet_scaling_2017}. Nevertheless, it maintains a leading market position \cite{MIV_dominance,MIV_dominance_2} because it is convenient \cite{kent2013secured} and relatively cheap for its users.  
 
A well-known answer to this problem is ride-pooling \cite{pooling_north_america,munich,data_driven_pooling}, i.e., combining routes of several passengers such that they can be served by one vehicle \cite{merlin2019transportation}. This is done most efficiently by line services, like trains, light rail, or the underground. A light rail car easily takes a hundred passengers or more, uses much less energy than the same number of MIV and contributes next to nothing to traffic congestion \cite{pitrzak}. Therefore, many large cities (like, e.g., Tokyo) rely heavily on transportation by line services \cite{pitrzak_cities,ferbrache_cities,kato}.
 
 They come, however, with a serious downside when compared to MIV. With the latter, users can freely choose the starting time and location as well as the destination. This is not possible for line services, which must follow fixed schedules and fixed routes \cite{Alam2015InvestigatingTD}. Users thus may have to walk significant distances to and from stations, and need to know the schedules of the involved lines. Demand-responsive ride-pooling (DRRP) \cite{herminghaus_mean_2019} services try to address this problem by deploying a large number of shuttles which pick up and drop off users at the desired  locations. This requires a central facility which collects travel requests, along with a powerful algorithm which combines these requests into appropriate routes of the shuttles \cite{alonso-mora_-demand_2017}. In such systems, users necessarily experience some detour \cite{herminghaus_mean_2019,lobel_detours_2020} with respect to the direct route they could have taken via MIV. This trade-off \cite{daganzo2020analysis} severely limits the achievable pooling efficiency to well below ten passengers per vehicle \cite{munich}. Hence while DRRP is more attractive than line services because it provides door-to-door transport, its pooling efficiency is intrinsically much inferior. 
 
In the present paper, we investigate a bi-modal public transit system, which consists of a combination of both transport modes. A line service, with fixed routes and schedule, shall coexist with a fleet of shuttles which pick-up users and bring them either to and from line service stations, or serve shorter-distance requests directly. This provides  both door-to-door transport by virtue of the shuttles and a large average pooling efficiency due to the involvement of line service vehicles. While there have been quite a number of studies on ride-pooling systems before \cite{nora_scaling_2020,tachet_scaling_2017,salonen,sorge,santi_shareability_2014,vazifeh_addressing_2018,data_driven_pooling,heuristic_pooling_quality,storch_incentive-driven_2021,storch_symmetry,Lotze_2022}, the combination of different transport modes has so far been only scarcely addressed and remains to be explored. We identify the relevant parameters which need to be controlled to optimize such combined system, aiming at a fully scalable description. We explore what performance may be achieved in terms of energy consumption, traffic volume, and quality of service.
\section{Definition of the system}

\subsection{User environment}

For the sake of conciseness and simplicity, we consider a planar area uniformly populated at density $E$ with potential users of the public transit system under study. Users are assumed to place transit requests in an uncorrelated fashion, each consisting of a desired pick-up ($\mathcal{P}$) and drop-off ($\mathcal{D}$) location, at an average rate $\nu$ per passenger. Requested travel distances $d = \overline{\mathcal{P} \mathcal{D}}$ are assumed to follow a certain distribution, $p(d)$, with mean $D$ \cite{herminghaus_mean_2019}.

For a transparent discussion, it is useful to introduce dimensionless parameters characterizing the system under study.  By combining the intrinsic length scale $D$  with a characteristic road vehicle velocity, $\shuttlespeed$, we obtain an intrinsic time scale, $t_0 = D/\shuttlespeed$. This is the average time a travel request would need to be completed by MIV.  The demand of transport within the system can then be characterized by the dimensionless parameter $\Lambda= E\nu D^3 / v_0\footnote{Average number of incoming requests in an area $D^2$ in time $t_0$.} $, which can reach well beyond $10^4$ in a densely populated area. Tab.~\ref{tab:demands} provides a few typical parameters encountered in real systems for reference. Note that $\tilde{\ell} = \ell/D = \sqrt{m}/D$, where $m$ is the average area enclosed by surrounding rail (line) services, and $\ell$  is the spacing of line service routes (see below, Fig.~\ref{fig:grid1}).

\begin{table*}
\centering
\begin{tabular}{lccccccccr} 
\toprule
  city/district & type & $E$                &  $D$  &$\shuttlespeed$ &$m$ & $\tilde{\ell}$ & $\Lambda$ & $\mathcal{Q}$ & ref. \\ 
&                &$[\si{km^{-2}}]$  &$[\si{km}]$&$[\si{km/h}]$ & $[\si{km^2}]$ &    &  &  & \\
\midrule

New York City  & dense urban & $1.  1 \cdot 10^4$  & $4.99$ &  $11.3$      & $2.0$ & $0.28$ & $1.5 \cdot 10^4$ & $0.33$ & \cite{herminghaus_mean_2019,NYC-census, NYC-taxi, NYC-speed}\\
Berlin         & urban       & $4.1 \cdot 10^3$    & $5.90$ & $19.8$       & $3.6$ & $0.32$ & $5.0 \cdot 10^3$ & $0.64$ & \cite{BER-census, mobinstaedten}\\
Ruhr (north)   & urban       & $3.6 \cdot 10^3$    & $15.7$ & $44.9$       & $94$  & $0.62$ & $3.6 \cdot 10^4$ & $0.34$ & \cite{Wirtschaftsdienst2018}\\
Emsland        & rural       & $1.1 \cdot 10^2$    & $16.7$ & $58.7$       & $1200$ & $2.1$ & $1.0 \cdot 10^3$ & $0.35$ & \cite{EMS-census, IAB2018}\\
\bottomrule
\end{tabular}
\captionsetup{textfont=md}
\caption{\textbf{City data.} Typical values of population density $E$, average traveled distance $D$, speeds of shuttles $\shuttlespeed$, as well as the resulting dimensionless demand $\Lambda = D^3 E \nu/\shuttlespeed$, service quality $\quality$ (see Eq.~\ref{eq:quality_word}) and dimensionless mesh size $\tilde{\ell}$, for a few selected areas. $\tilde{\ell} = \sqrt{m}/D$, where  $m$ is the average area enclosed by surrounding rail services. We assume $\nu = 2/17 \,\si{\per\hour}$, i.e., two trips per day per user given a time of service of $17\,\si{\hour}$ per day. Road vehicle velocities for Ruhr (north) and Emsland, as well as data for $\mathcal{Q}$, have been obtained by averaging Google navigator data over many relations randomly chosen within the respective area.}
\label{tab:demands}
\end{table*}

\subsection{Model system geometry}
 
\begin{figure*}
    \centering
    \begin{subfigure}[b]{0.44\linewidth}
    \centering
    \includegraphics[width=\columnwidth]{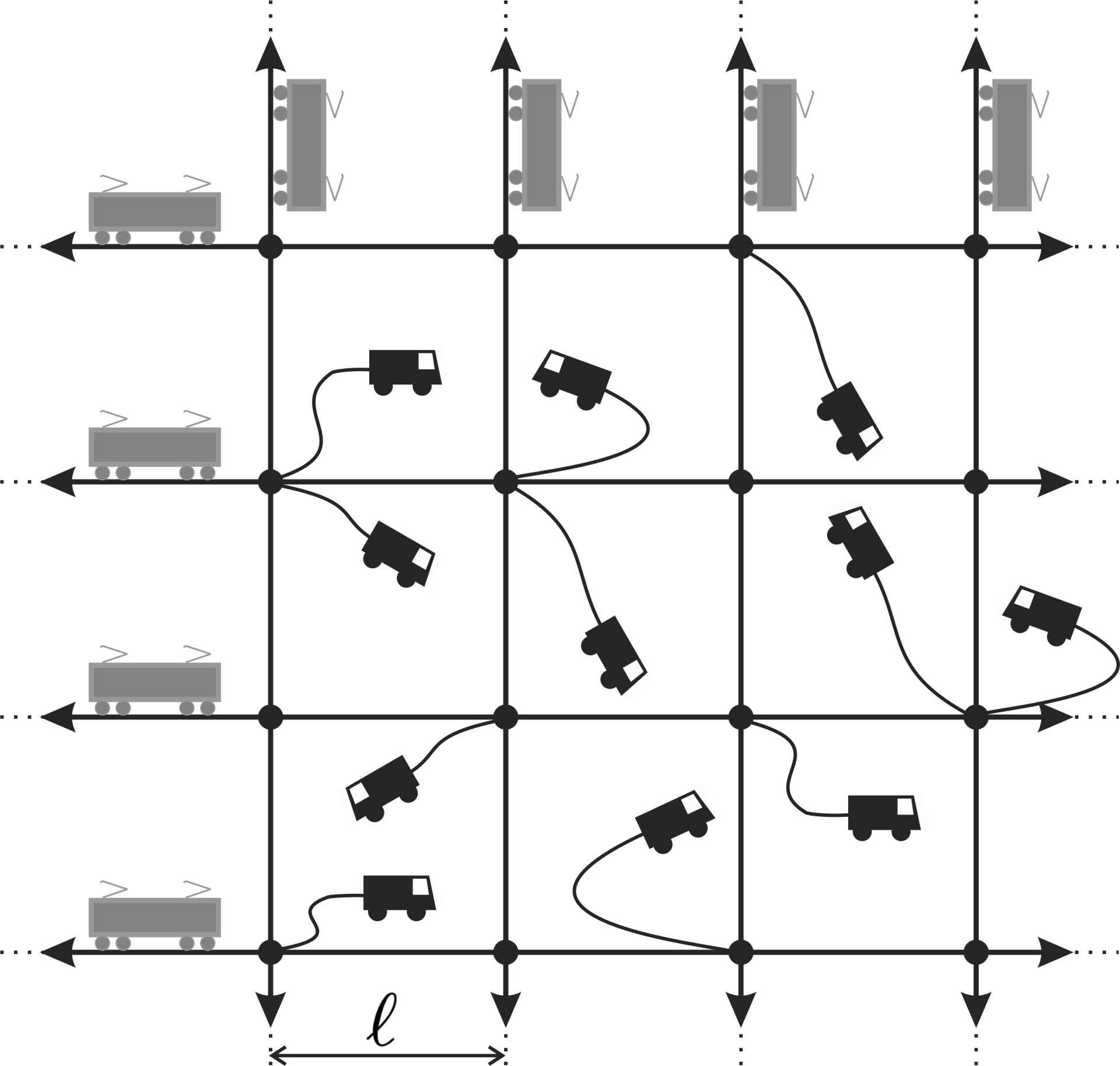}
    \caption{}
    \label{fig:grid1}
    \end{subfigure}
    \hspace{5mm}
    \begin{subfigure}[b]{0.4\linewidth}
    \centering
    \includegraphics[width=\columnwidth]{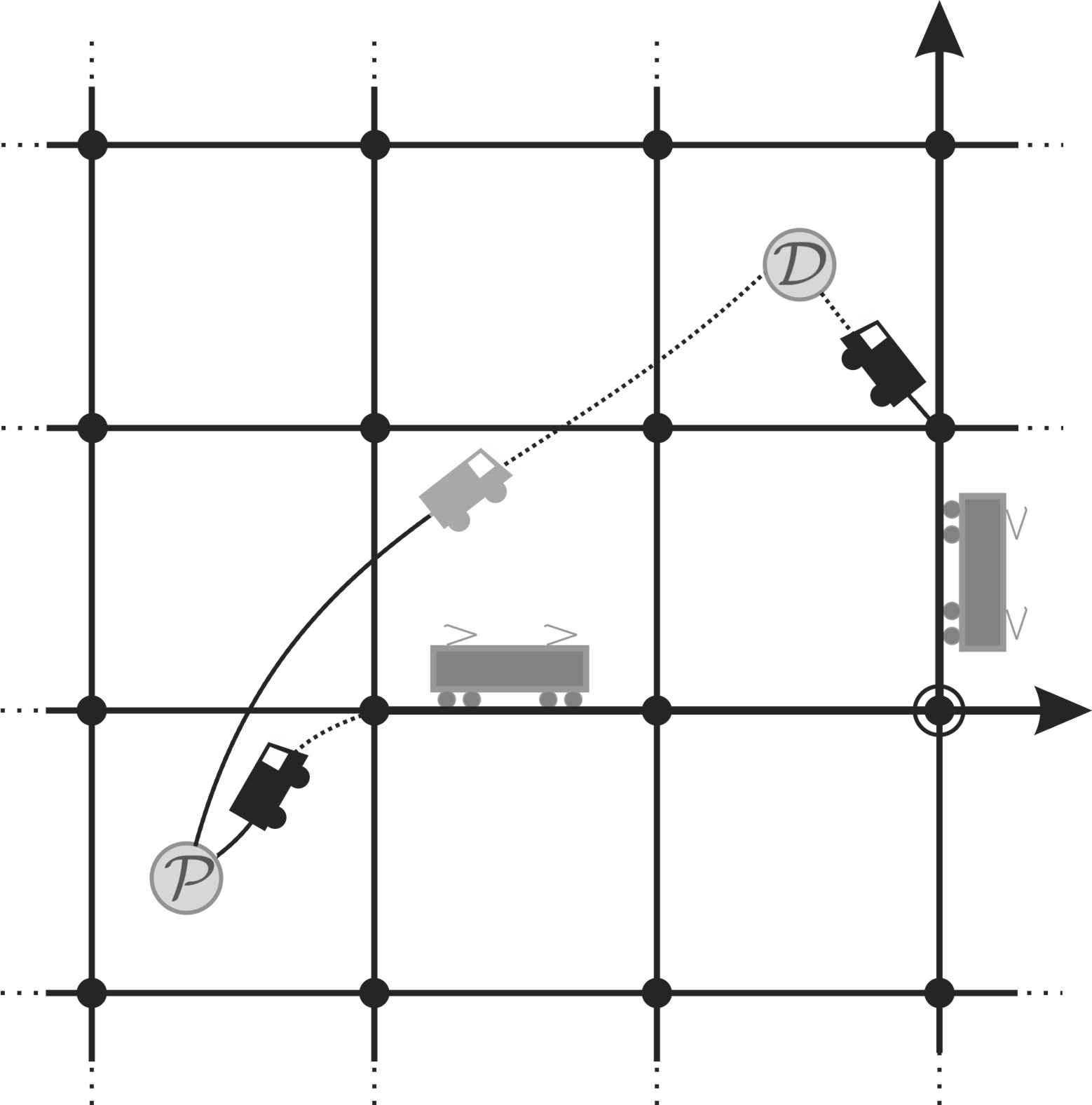}
    \caption{}
    \label{fig:grid2}
    \end{subfigure}
     \caption{\textbf{Bi-modal transport network on a square grid.} \textbf{(a)} A bi-modal network with trains (grey vehicles) operating along the solid lines. Shuttles (black vehicles) are used as a feeder service to carry people to and from the train stations (black dots at intersection points) separated by distance $\ell$. Trains operate periodically at a frequency $\mu$, with vehicle seating capacity $k$.
     \textbf{(b)} Two alternative ways to serve a transport request from $\mathcal{P}$ (pick-up) to $\mathcal{D}$ (drop-off). Bi-modal transport involves a shuttle ride from $\mathcal{P}$ to the train station, transport by train (arrows, here with one change (circle)), and another shuttle ride from the train station to $\mathcal{D}$. Uni-modal transport service is a direct shuttle (grey) ride from $\mathcal{P}$ to $\mathcal{D}$. A major task of the system is to appropriately decide which of these two types of transport services to choose.}
\end{figure*}

For an overarching systematic study, it is useful to consider an idealized model geometry (see~Fig.~\ref{fig:grid1}). We  assume that transport occurs via DRRP shuttle service, combined with a square grid of railways on which transport occurs via trains. The connection points (train stations) between the two subsystems lie at all railway intersections and are spaced with a lattice constant $\ell$ (cf.~Fig.~\ref{fig:grid1}).
The transit system is further characterized by a shuttle density $B$ in the plane and a train frequency $\mu$ at all train stations, with trains having a seating capacity $k$.
Shuttles and trains  move with velocities $\shuttlespeed$ and $\trainspeed$, respectively. They require energy $\shuttleEnergy$ and $\trainEnergy$, respectively, per unit distance of travel. 

The main goal of the bi-modal system under consideration is to provide high quality (i.e., rapid) door-to-door transportation service at minimal energy consumption, thereby minimal carbon emission. 
To reach this goal, the provider of bi-modal transit may vary certain parameters of operation.
We will first introduce these control parameters in Sec.~\ref{sec:parameters} and then derive expressions for the system's service quality and energy consumption as functions of these parameters in Sec.~\ref{sec:objectives}.

\section{Parameters of operation}
\label{sec:parameters}

\subsection{Choosing the type of transport service}

\begin{figure*}[ht]
    \centering
    \includegraphics[width=\linewidth]{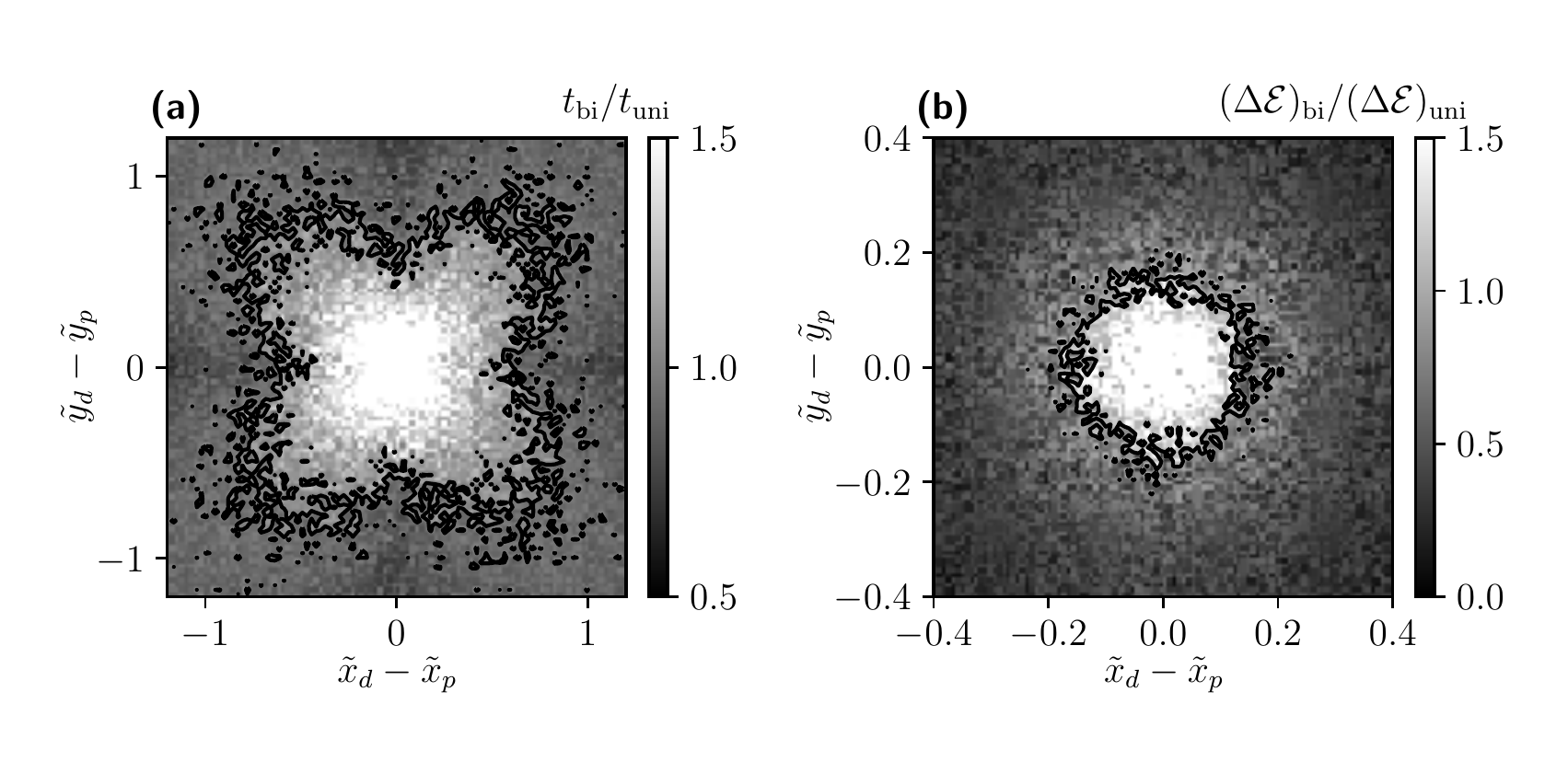}
    \caption{\textbf{Choosing the type of transport service.} Relative characteristics of either bi-modal (shuttle-train(s)-shuttle) or uni-modal (just shuttle) service, in the plane spanned by the individual trip vector from pick-up $\mathcal{P} = (x_{p},y_{p})$ to drop-off $\mathcal{D} = (x_{d},y_{d})$. \  \textbf{(a)} Bi-modal travel time, $t_\textrm{bi}$, divided by uni-modal travel time, $t_\textrm{uni}$. The black curve represents the contour line where both are equal. Requests outside this region are served faster with bi-modal transportation. \  \textbf{(b)} Increment in total energy consumption if a new user is served by bi-modal transportation, $(\Delta \mathcal{E})_\mathrm{bi}$ divided by the increment in total energy consumption if the same user is served by uni-modal transportation, $(\Delta \mathcal{E})_{\mathrm{uni}}$. The black curve represents the contour line where both are equal, i.e., from the perspective of energy consumption both types of transport service are equivalent. Requests outside the white region lead to lower energy consumption if served by bi-modal transportation. See \SuppInf for details.}
    \label{fig:cutoff_motivation}
\end{figure*}

A single user in the model system may either be transported by uni-modal service, i.e., by shuttle (DRRP) only, or by bi-modal service, i.e., be brought from $\mathcal{P} = (x_{p},y_{p})$ to the nearest train station by means of a shuttle, followed by a train journey, which is again followed by a shuttle journey to $\mathcal{D} = (x_{d},y_{d})$ (see Fig. \ref{fig:grid2}). Aside from assembling the routes of the shuttles such as to optimize pooling efficiency, one central task of the dispatcher system will be to decide, for each individual request $(\mathcal{P}, \mathcal{D})$, whether the desired door-to-door service should be completed uni-modally or bi-modally.
 
If only user convenience were considered relevant, one would  just need to calculate which type of transport service requires less time for completing the transit, and then to choose that one.
This requires knowledge of the parameters $v_0$, $v_{\textrm{train}}$, and the frequency of line service, $\mu$.
The latter can be assumed to be just sufficient to carry the bi-modal passenger load.
Its derivation will be discussed further below (2.\ref{subsec:mu}).
By choosing random transport requests in the plane, while sampling distances from the probability distribution $p(d)$ of travel distances, we can compile a histogram of relative travel times, as shown in Fig.~\ref{fig:cutoff_motivation}a.
It displays the resulting ratio $t_{\textrm{bi}}/t_{\textrm{uni}}$\footnote{Subsection 3.\ref{sub:quality} displays the mathematical expressions of $t_{\textrm{bi}}$, $t_{\textrm{uni}}$.} as a scatter heat map in the plane spanned by the vector $\overline{\mathcal{P}\mathcal{D}}$. 
Requests corresponding to the area within the black curve (contour curve of $t_{\textrm{bi}}/t_{\textrm{uni}} = 1$) would then be served uni-modally by a single shuttle services, while for all others, the dispatcher would offer bi-modal transport service. 

In order to choose, for an incoming request $(\mathcal{P},\mathcal{D})$, the type of transport service  which consumes the least incremental amount of energy, we have to compute the energy increment $(\Delta \mathcal{E})_\mathrm{bi}$ needed for bi-modal transport, and compare it to the energy increment $(\Delta \mathcal{E})_\mathrm{uni}$ assuming direct transport via a single shuttle.
This ratio of energy increments is equal to the ratio of driven distances by the shuttles for each type of service (see also Fig.~\ref{fig:grid2}). We assume that a single request does not alter the line service frequency, i.e, the energy consumption of the line service does not change. 

In analogy to the travel times shown in Fig.~\ref{fig:cutoff_motivation}a, the ratio of the energy consumption increments (i.e., traveled distances by shuttles) is depicted in Fig.~\ref{fig:cutoff_motivation}b.
Again, we see that while for small requested distances uni-modal service is advisable, bi-modal service should be preferred for larger distances, corresponding to the area outside the black contour curve. 

Comparison of Figs.~\ref{fig:cutoff_motivation}a and \ref{fig:cutoff_motivation}b  reveals that there is a significant range of distances which lie outside the solid curve in Fig.~\ref{fig:cutoff_motivation}b, but still well inside the curve depicted in in Fig.~\ref{fig:cutoff_motivation}a. This shows that we may have to deal with conflicting objectives for quite a number of incoming transport requests. The notion of optimality then depends upon the relative valuation of energy consumption and service quality. As a generally accepted way of dealing with conflicting objectives, we will tackle this problem by means of Pareto fronts \cite{debreu1959,greenwald1986,magill2002} further below (4.\ref{subsec:Pareto}). 
 
While the plot in Fig.~\ref{fig:cutoff_motivation}b represents a rather isotropic structure, we encounter a shamrock-like shape in Fig.~\ref{fig:cutoff_motivation}a. This reflects the orthogonal geometry of our model line service system (Fig.~\ref{fig:grid1}). In a real situation, the geometry will in general not be this simple. Instead, the directions at which rails are installed will vary from one station to another. We thus expect a structure like the `shamrock' to be less pronounced in reality, if discernible at all.  Hence although a perfectly isotropic structure may not be expected, the anisotropy will certainly be less pronounced. We assume that it will be a reasonable approximation to consider the contour line of service times as `sufficiently' circular.
Therefore we consider henceforth only the requested travel distance, $d = \vert \overline{\mathcal{P}\mathcal{D}} \vert$, as the discriminating parameter for the choice of type of transport service, irrespective of its direction. The task of the dispatcher will then be to determine a proper cutoff distance, $\dcut$, such that for $d > \dcut$, bi-modal service is offered, while for $d \le \dcut$, the system will provide uni-modal service, by shuttle only. 
 
Note that the above approximation provides a lower bound of the performance achievable. In a real system, the type of transport service may be decided upon the true expected travel times  and energy consumption, for which data will be available with ever improving quality over time.   
 
\subsection{Choice of line service frequency}\label{subsec:mu}

It is clear that the capacity $k$ and frequency $\mu$ of the line service must be sufficient to carry the flux of shuttle passengers towards and from the train stations.  The total number of requests emanating in unit time in an area of $\ell^{2}$ around a train station is $\nu E \ell^{2}$.
Out of these, only a fraction $\fbi = \fbi(\dcut) = \int_{\dcut}^{\infty} p(d)\ \textrm{d}d$
is served by bi-modal transportation. However, trains are also occupied by passengers from previous stations. If $D_{\textrm{train}}$ is the average distance that users travel on trains, then a user travels $D_{\textrm{train}}/\ell$ stations on train on average. Therefore, the total number of users to be transported at this station per unit time is
\begin{equation}
    \mathcal{J}_{\textrm{in}} = \nu E \ell^{2} \fbi\frac{D_{\textrm{train}}}{\ell}\,.
\end{equation}
We find that $D_\textrm{train} = \frac{4}{\pi}\dgdc$\footnote{Averaging the $1$-norm $\Vert \overline{\mathcal{P}\mathcal{D}} \Vert_1$ over distances and orientations.}, with $\dgdc$ the mean of requested distances larger than $\dcut$.

A similar relation holds for the number of users per unit time that can be transported by trains arriving at one train station (with frequency $\mu_0$ and going into four directions), namely
\begin{equation}
    \mathcal{J}_\textrm{out} = 4\cdot{\mu_0}\cdot k \,.
\end{equation}
Balancing $\mathcal{J}_\textrm{in}$ with $\mathcal{J}_\textrm{out}$, we obtain
\begin{equation} \label{eq:SD_derv_2}
\tilde \mu_0 =   \frac{\Lambda \tilde\ell}{\pi k} \dgdctilde \fbi\,
\end{equation}
for the minimum frequency required to carry all passengers conveyed by the shuttles. The $\ \tilde{ }\ $ indicates quantities non-dimensionalized via division by the respective unit, i.e., $D$ or $t_0$.
We refer to  Eq.~\ref{eq:SD_derv_2} as passenger flux balance.

If we allow trains to operate at a frequency $\tilde \mu$ larger than the minimum required frequency $\tilde \mu_0$, the train occupancy is given by $\bm{\alpha}=\tilde \mu_0/\tilde \mu \in [0,1]$.
As this can be adjusted within some range when operating the line service, $\bm{\alpha}$ provides an additional free variable in system operation.

\section{Objectives of operation}
\label{sec:objectives}

\subsection{Service quality}
\label{sub:quality}
 
We define the service quality as the ratio between average travel time by MIV and by the bi-modal system:
\begin{equation}
\label{eq:quality_word}
\quality = \frac{t_0}{\funi \cdot t_{\mathrm{uni}} + \fbi \cdot t_{\mathrm{bi}}}\,.
\end{equation}
For assessing the overall quality of service, suitable averaging has to be applied. 
Transportation by shuttles is always assumed to be delayed with respect to MIV by a waiting time, which we assume (on average) to be of order one half the direct travel time, $t_0/2$ (see \SuppInf~for motivation).
 
The average time taken to serve a request in a bi-modal system (i.e., the denominator of $\mathcal{Q}$ in Eq.~\ref{eq:quality_word}) is then
\begin{equation}
\begin{aligned}
    t_0 \quality^{-1} = &(1-\fbi) \cdot \underbrace{\left(\frac{t_0}{2}+\frac{\delta\dldc}{\shuttlespeed}\right)}_{t_\mathrm{uni}} + \fbi \cdot \underbrace{\left(t_0 +\frac{2\beta\ell\delta}{\shuttlespeed} +  \frac{1}{\mu}+\frac{4}{\pi }\frac{\dgdc}{\trainspeed}\right)}_{t_\mathrm{bi}}\,,
\end{aligned}
\end{equation}
where $\dldc$ represents the mean of all requested distances less than $\dcut$ (i.e., served uni-modally) and $\delta$ is the average detour incurred by the shuttles due to the necessity of pooling several different transport requests into one vehicle route. For the expected detour with a shuttle we set $\delta = 1.5$ (see \SuppInf~for details).
For the second term ($t_\mathrm{bi}$), $t_0$ is the total average waiting time for two shuttle trips (to and from the station), $1/\mu$ is the average waiting time for two train rides (usually there is a change involved), $\beta \ell$ is the average distance of a randomly chosen point from the next train station, with a geometrical constant $\beta \approx 0.383$\footnote{A simple calculation shows that $\beta = \frac{1}{6}(\sqrt{2}+\log(1+\sqrt{2})) = 0.383$.}, and $4\pi^{-1}\dgdc$ is the average distance traveled on trains.
The effective train velocity, $\trainspeed$, depends on the inter-station distance $\ell$ and is modeled based on train vehicle data (see~\SuppInf~for details). 

If we use $D$, $t_0$, and $v_0$ as units for length, time, and velocity, respectively, we can write:
\begin{equation}
\label{eq:quality_math}
    \begin{aligned}
         \quality^{-1} =  (1-\fbi)\cdot \left(\frac{1}{2} + \delta \dldctilde \right) + \fbi \cdot \left(  1 + 2\beta  \tilde{\ell}\delta + \frac{1}{\tilde{\mu}} 
         +  \frac{4}{\pi}\frac{\dgdctilde}{\tilde{v}_\textrm{train}}  \right)\,.
    \end{aligned}
\end{equation}

\subsection{Energy consumption}

In order to assess the efficiency of a transit system in terms of energy consumption, it is essential to consider the total distances over which passengers are being transported in the different vehicles involved (see~Eq.~\ref{eq:emission_word}). The bi-modal energy consumption can be written as
\begin{align}\label{eq:emission_word}
\mathcal{\emissions} \equiv \frac{\Delta_\mathrm{shuttle} \cdot \shuttleEnergy+\Delta_\mathrm{train} \cdot \trainEnergy }{\Delta_\mathrm{MIV} \cdot \carEnergy}\,,
\end{align}
where $\Delta_{\cdot}$ denotes the (mode-specific) total distance traveled in a unit cell of size $\ell^2$ per unit time, and $e_{\cdot}$ is the vehicle-specific energy consumption per unit distance. Note that this expression is already normalized with respect to the MIV energy consumption (denominator), as this is the door-to-door transportation system we intend to compare with. For $\emissions > 1$ ($<1$) energy requirement for bi-modal transportation is more (less) than for private cars serving the same requests.

For the simulations we will describe below, we use numbers found for frequently used transport vehicles. Specifically, we consider electric light rails with a maximum seating-capacity $k=100$ and $\trainEnergy = 9.72\ \si{kN}$ \cite{TREMOD} for the line service. For MIV we consider Diesel cars with $\carEnergy=2.47\ \si{kN}$ \cite{viz2021}. For the shuttle we choose  Mercedes Sprinter ($8.8$ liters of Diesel per $100\ \si{km}$  \cite{sprinter_energy}), thus $\shuttleEnergy = 3.28\ \si{kN}$.

\paragraph{Shuttles.} Both uni-modal (shuttle only) and bi-modal trips contribute to the total distance driven by shuttles per unit time due to requests from a unit cell of size $\ell^2$, hence
\begin{equation}\label{eq:d_shuttle}
    \Delta_\textrm{shuttle} = \frac{\nu E \ell^{2}}{\etabi} \left(\underbrace{\dldc\funi}_{\textrm{shuttle only}} + \underbrace{2\beta\ell\fbi}_{\textrm{two shuttle trips}}\right)\,,
\end{equation}
where $\eta$ is the DRRP pooling efficiency, which is the ratio of requested direct distance by the users and the driven distance by the shuttles (for MIV, $\eta = 1$).

In simulations of the uni-modal system (shuttles only), it has been observed that $\eta$ scales with demand $\Lambda$ roughly in an algebraic manner, $\eta (\Lambda) \propto \Lambda^{\gamma}$, with $\gamma\approx 0.12$ \cite{Muehle2022}. In a bi-modal system, however, some of the demand $\Lambda$ is directed towards trains, therefore, we need to compute an adjusted demand, $\lambdabi \equiv (E\nubi \Dbi^3)/\shuttlespeed$, considering shuttle trips only; $\nubi$ is the effective request frequency for shuttle trips and $\Dbi$ is the average distance of a shuttle trip. Bi-modal trips consist of two shuttle trips (from and to the station), therefore
\begin{align}\label{eq:eta_nu}
    \nu_{\mathrm{shuttle}} &= \underbrace{2 \nu \fbi}_\textrm{two shuttle trips} + \underbrace{\nu \funi}_\textrm{shuttles only} = \nu(1+\fbi)\,.
\end{align}
Similarly, the average requested distance for shuttle-borne trips involved in bi-modal transport is
\begin{align}\label{eq:eta_D}
    D_{\mathrm{shuttle}} &= \big(\underbrace{2 \beta \ell \fbi}_\textrm{two shuttle trips} + \underbrace{\dldc \funi }_\textrm{shuttles only}\big)/(2-\fbi)\,,
\end{align}
where $(2-F)$ is due to normalization. The bi-modal demand for shuttles is thus given by:
\begin{align}\label{eq:lmbd_shuttle}
    \Lambda_\mathrm{shuttle} &= (E \nu_\mathrm{shuttle} D_\mathrm{shuttle}^3)/\shuttlespeed \nonumber\\
    & = \Lambda \, (1 + \fbi)^{-2}(\funi \dldctilde + 2 \beta \tilde{\ell} \fbi)^3\,.
\end{align}

In simulations we observe a higher efficiency than suggested by $\eta \propto \lambdabi^{0.12}$ (see \SuppInf~for simulation data). We call this the \emph{common stop effect}, meaning that pooling gets more efficient because bi-modal requests are spatially correlated due to shared pick-up and drop-off locations, i.e., the train stations. We account for this effect via an empirical function $h(\fbi)$ ($1\le h \le 1.35$, see \SuppInf~for details). In particular, we set
\begin{align} 
\label{eq:eta_bi}
    \eta \equiv \Lambda_\mathrm{shuttle}^{0.12} \cdot h(\fbi)\,.
\end{align}

\paragraph{Line Service and MIV.} Trains are recurrent every $1/\mu$ time units.
Therefore, the cumulative distance driven by all trains in a unit cell of side length $\ell$ per unit time is
\begin{equation}\label{eq:d_train}
    \Delta_\textrm{train} = 4\cdot\mu\cdot\ell\,.
\end{equation}

There is a multiplicative factor of 4 because trains go in four directions at every train station.
The total distance driven via MIV for requests from the unit cell amounts to
\begin{equation}\label{eq:d_MIV}
    \Delta_\textrm{MIV} = \nu E \ell^{2} D.
\end{equation}

Replacing $\Delta_\textrm{shuttle}$, $\Delta_\textrm{train}$,  and  $\Delta_\textrm{MIV}$ in Eq.~\ref{eq:emission_word} from Eq.~\ref{eq:d_shuttle}, \ref{eq:d_train}, and \ref{eq:d_MIV} we obtain the final expression for the energy consumption of bi-modal transit normalized with respect to MIV:
\begin{equation}\label{eq:emission_math}
\begin{aligned}
    \emissions=  \underbrace{\eta^{-1}\left( \dldctilde\funi + 2\beta\tilde{\ell} \fbi \right)\cdot\frac{\shuttleEnergy}{\carEnergy}}_\textrm{shuttles}  + \underbrace{\frac{4\tilde{\mu}}{\Lambda \tilde{\ell}}\cdot\frac{\trainEnergy}{\carEnergy}\,.}_\textrm{train}
\end{aligned}
\end{equation}

\section{Results}
 
We now analyze how the objectives, i.e., energy consumption (Eqs.~\ref{eq:emission_word},  \ref{eq:emission_math}) and quality (Eqs.~\ref{eq:quality_word}, \ref{eq:quality_math}), can be optimized by choice of parameters of operation, i.e., cutoff distance $\dcut$ and train occupancy $\trainocc$, under different `external' conditions, $\Lambda$ and $\tilde{\ell}$.
Notice that the two control parameters, $\trainocc$ and $\dcut$, enter the objectives, $\quality$ and $\emissions$, via $\langle \tilde{d} \rangle_{\tilde d \lessgtr  \tilde \dcut}$, $F(\dcut)$, and $\tilde \mu= \tilde \mu_0(\dcut)/\trainocc $ (Eq.~\ref{eq:SD_derv_2}).

\begin{figure}[ht]
\centering
\includegraphics{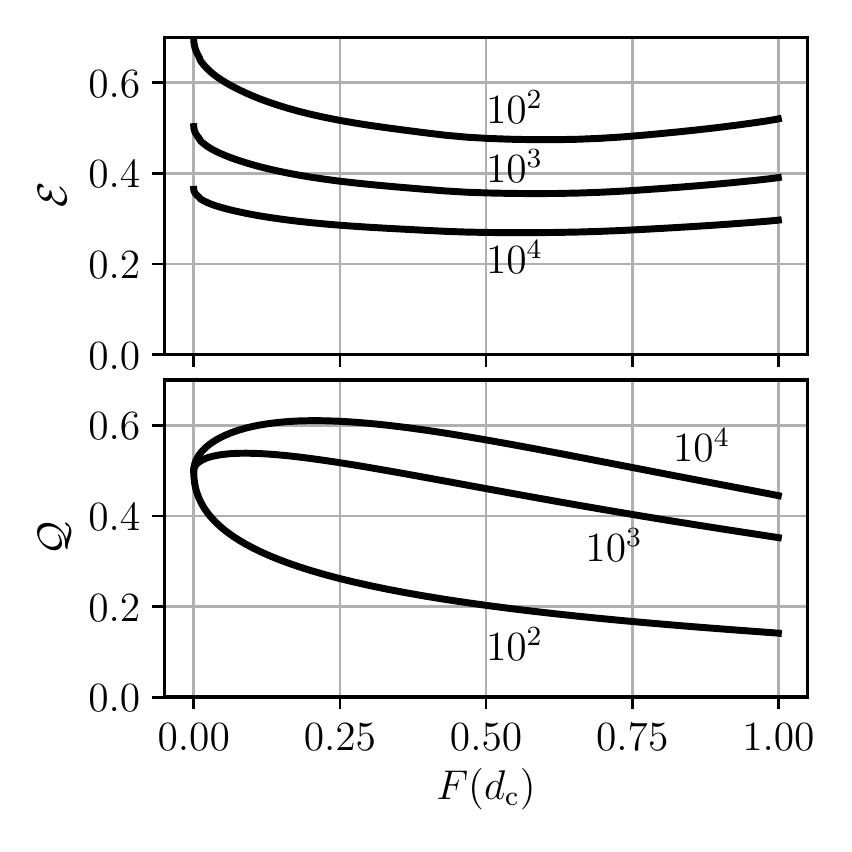}
\caption{\textbf{Bi-modal performance characteristics.} Energy consumption $\emissions$ and service quality $\quality$ for bi-modal transport, normalized with respect to MIV, as a function of the bi-modal fraction $\fbi(\dcut)$, for three different values of demand $\Lambda=\{10^2,10^3,10^4\}$. All data for $\tilde{\ell} = 0.8$ and fully-occupied trains, $\trainocc=1$.}
\label{fig:pareto-intro}
\end{figure}

In Fig.~\ref{fig:pareto-intro}, energy consumption, $\emissions$, and service quality, $\quality$, for the combined  system are shown as a function of the share of bi-modal transport $\fbi(\dcut)$ at $\tilde{\ell} = 0.8$ for three different values of dimensionless demand, $\Lambda$.
Trains are operated at full occupancy, i.e., $\trainocc=1$.
The general trend of reduction of energy consumption with increasing demand and involvement of line services is obvious from the data for $\mathcal{E}$.
We encounter a minimum of energy consumption at around $\fbi \approx 0.6$ for all values of $\Lambda$ investigated.
Energy consumption can be less than $30\%$ of MIV for sufficiently large (but realistic, see Tab.~\ref{tab:demands}) demand.
For service quality, we find that typical values are around one half the service quality of MIV, i.e., about twice the travel time.
This is customary for public transport systems \cite{SALONEN2013143,liao2020disparities} and generally well accepted by users. Note that our data for service quality represent a safe lower bound, as the (sometimes quite substantial) time required for parking spot search \cite{fulman2021approximation,CHANIOTAKIS2015228} is neglected here in $t_0$.
While for low $\Lambda$ the involvement of line service seems to generally increase travel times (i.e., reduce service quality), we find an optimum at $\fbi \approx 0.25$ for large $\Lambda$.
The primary message from Fig.~\ref{fig:pareto-intro}, however, is that minimizing energy consumption and maximizing service quality cannot be simultaneously achieved.    

\subsection{Pareto fronts in energy consumption and service quality}\label{subsec:Pareto}

\begin{figure*}
    \centering
    \includegraphics{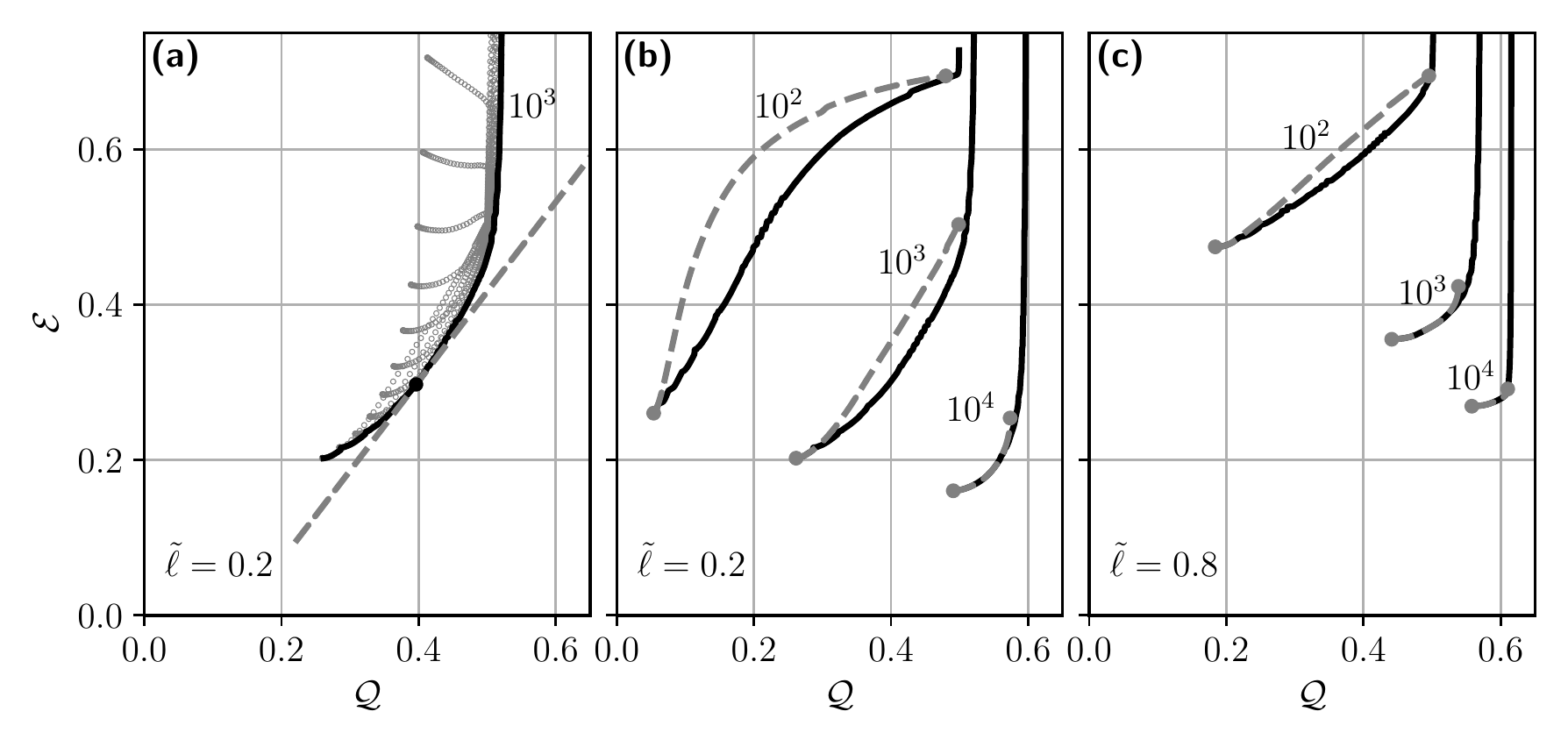}
    \caption{\textbf{Emergence of Pareto fronts and effects of train occupancy}. \ \textbf{(a)} Grey circles: admissible data for full variation of $\trainocc$ and $\dcut$, at $\tilde{\ell}=0.2$ and $\Lambda=10^{3}$.
    Black curve: Pareto front, i.e., the boundary of the full data set towards optimality (low $\mathcal{E}$ and high $\mathcal{Q}$). The slope of the dashed tangent to the Pareto front represents the ratio of valuations (see text).  \ ~\textbf{(b)}. Black curves: Pareto fronts for variable train occupancy $\trainocc<1$ at $\Lambda=\{10^{2},10^{3}$,$10^{4}\}$ and $\tilde{\ell}=0.2$.
    Grey dashed curves: degenerate Pareto fronts obtained at full train occupancy ($\trainocc=1$) at corresponding values of $\Lambda$.
    Grey circles mark the ends of these fronts which are determined by minimum achievable energy consumption and maximum achievable service quality, specific to $\Lambda$ and $\tilde{\ell}$.~\textbf{(c)} Same as (b) but for $\tilde{\ell} = 0.8$.
    }
    \label{fig:noSD}
\end{figure*}

A tuple of parameter values, in our case $(\emissions, \quality)$, is called Pareto optimal if none of the parameters (or objectives) can be further optimized without compromising on at least one of the others. The set of all such tuples of parameters is called the Pareto front. 
To illustrate this concept, in Fig.~\ref{fig:noSD}a, we show all values $\{\emissions(\dcut,\trainocc), \quality(\dcut,\trainocc)\}$ obtained for different values of $\dcut$ and $\trainocc$ as grey dots.
The solid black line represents the Pareto-optimal subset, i.e., the Pareto front.
 
In order to choose the truly optimal point on the Pareto front, one needs to define the relative valuation of the objectives, $\mathcal{E}$ and $\mathcal{Q}$. In other words, the authorities operating the system have to decide how much reduction in service quality they (i.e., the users) would be willing to accept for how much savings in energy. The ratio of these valuations is then expressed in the slope of the dashed line in Fig.~\ref{fig:noSD}a, which is a tangent to the Pareto front. The point where it touches the front (black dot) represents the optimal set of parameters, under the given valuation.    
     
For our analysis, we fix the train capacity to $k=100$, and choose a representative set of values for line service mesh size  $\tilde{\ell} = \{0.2,0.4,0.8\}$ and demand $\Lambda = \{10^2,10^3,10^4\}$, corresponding  to a typical parameter range encountered in real systems  (see~Tab.~\ref{tab:demands}). Note that fixing $k$ does not reduce the generality of our study, because a different $k$ can be compensated for by properly readjusting $\mu_0$ (see Eq.~\ref{eq:SD_derv_2}). 

Fig.~\ref{fig:noSD}b demonstrates the effect of $\Lambda$ on the overall performance of the system. Pareto fronts are shown in black. We see that for typical values of $\Lambda$, as listed in Tab.~\ref{tab:demands}, energy consumption reaches down to well below 40\% (even below 20\%) of MIV for large values of $\Lambda$. At the same time, quality levels are comparable to, mostly even in excess of, what is found in existing public transport  in terms of transit time (see~Tab.~\ref{tab:demands}). Note, however, that our system even provides on-demand door-to-door service, comparable to MIV. 
 
The dashed grey curves indicate the subset of data for $\trainocc=1$. We will henceforth call them degenerate Pareto fronts, as they correspond to the variation of only one parameter. They are lying, slightly but consistently, above the Pareto fronts. This indicates that by reducing train occupancy below its maximum ($\trainocc<1$), one can enhance the overall performance of the system. This is because the time spent waiting for trains, which is proportional to $1/\mu$, is smaller when trains need to be operated more frequently, to compensate for smaller occupancy. As this waiting time is inversely proportional to both $\Lambda$ and $\tilde \ell$ (see Eq.~\ref{eq:SD_derv_2}), this effect is more pronounced for small $\Lambda$ and $\tilde{\ell}$. This is found to overcompensate the increase in energy consumption due to higher operation frequency. It is qualitatively corroborated as well in Fig.~\ref{fig:noSD}c which shows corresponding data for large mesh size ($\tilde \ell = 0.8$). At the resolution of the figure, the Pareto fronts (black) and their degenerate partner (dashed grey) are distinguishable only for smallest values of $\Lambda$. Hence for typical values of $\Lambda$ and $\tilde \ell$ it appears sufficient to discuss the degenerate Pareto fronts, which only need one parameter $(\dcut)$ to be varied. We keep in mind that the true Pareto optimum will still be superior.

\begin{figure*}[ht]
\centering
\includegraphics{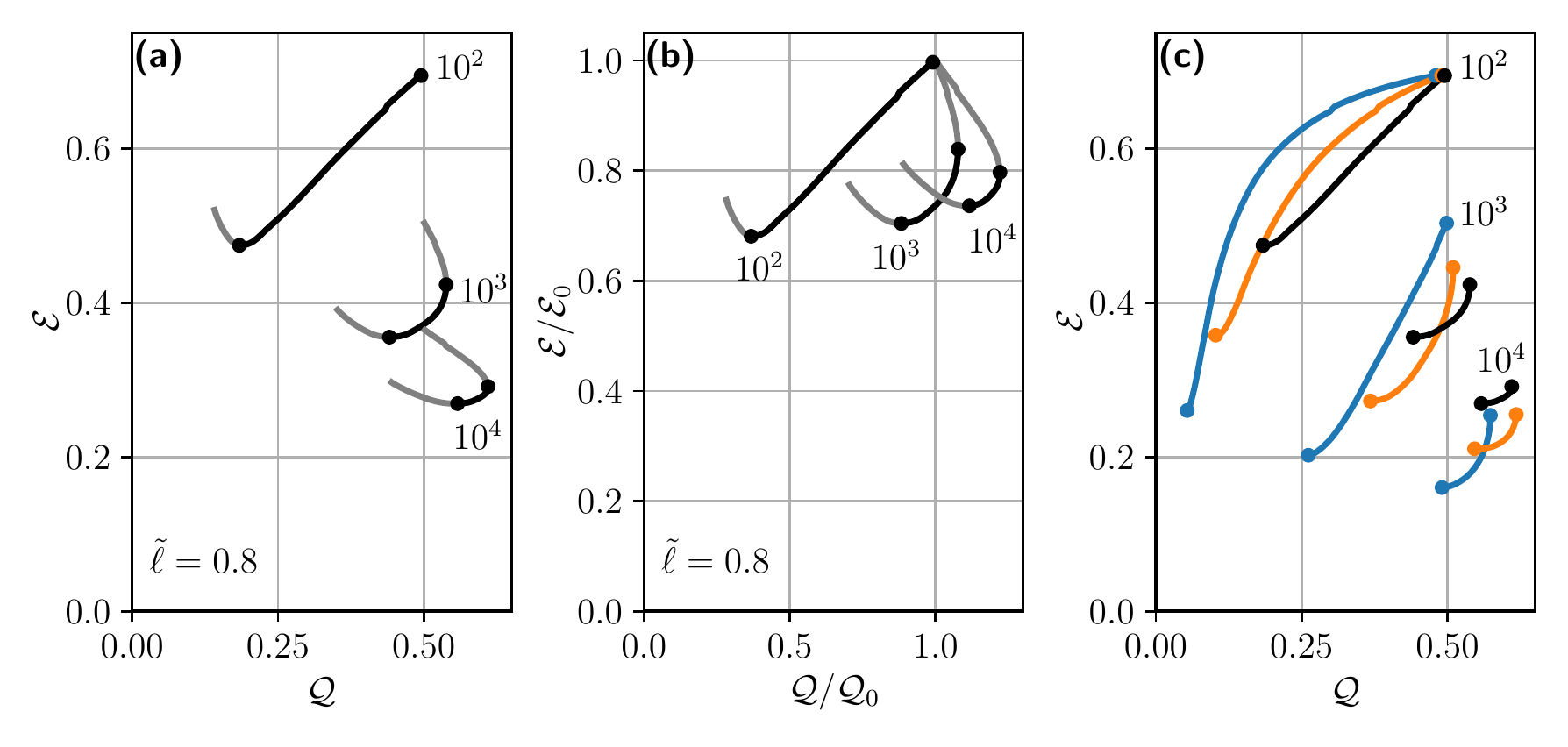}
\caption{\textbf{Bi-modal performance with fully-occupied trains. (a)} Black curves show degenerate Pareto fronts for fully-occupied trains ($\trainocc = 1$) for varying demands $\Lambda=\{10^2,10^3,10^4\}$ shown as annotations and $\tilde{\ell} = 0.8$.
Black circles mark the end points of the Pareto fronts which are determined by the minimum achievable energy consumption and the maximum achievable service quality.
Grey curves show the entire data, i.e., not only the Pareto-optimal set, but all admissible values with $\dcut$ as the control parameter.
\textbf{(b)} Data as in (a), but normalized with respect to the performance, $(\mathcal Q_0, \mathcal E_0)$, of the uni-modal system (shuttles only).  \textbf{(c)} Degenerate Pareto fronts as in (a), but for $\tilde{\ell} = \{0.2,0.4,0.8\}$ in blue, orange, and black, respectively.
}
\label{fig:pareto-intro-miv}
\end{figure*}

Degenerate Pareto fronts are shown in Fig.~\ref{fig:pareto-intro-miv} in various presentations. Fig.~\ref{fig:pareto-intro-miv}a has basically the same information as Fig.~\ref{fig:noSD}c, but here we show the full data set (grey), where the black curves are only the degenerate Pareto fronts. They terminate wherever the tangent becomes either vertical or horizontal (black dots), thus offsetting any tradeoff between the objectives. 
 
Fig.~\ref{fig:pareto-intro-miv}b shows these data normalized with respect to the performance of the uni-modal (shuttles only) DRRP system, represented by $\mathcal{E}_0$ and $\mathcal{Q}_0$. Clearly in relevant parameter regions the combined system outperforms the uni-modal system in both energy consumption ($\mathcal{E} < \mathcal{E}_0$) and service quality ($\mathcal{Q} > \mathcal{Q}_0$). 
  
The effect of inter-station distance (mesh size) on the (degenerate) Pareto fronts is explored in Fig.~\ref{fig:pareto-intro-miv}c. We see that a dense network of rails (small $\meshsize$, blue fronts) achieves the best results concerning energy consumption, reaching down to below 20\% of MIV, but compromises on achievable service quality. For admissible quality $\quality\approx 0.5$, sparse train networks (black fronts) are Pareto-optimal for low demands. For larger demands, denser train networks (orange and blue fronts) are advantageous. 
 
However, it is remarkable that the overall position of the Pareto fronts in the plane spanned by $\mathcal{Q}$ and $\mathcal{E}$ does not vary dramatically with mesh size, as the position on the front at which the system is operated is largely at the discretion of the operator. This suggests that the density of currently installed rail track systems might already be well suited for deploying a bi-modal on-demand transport systems of the kind we have studied.

\subsection{Traffic volume}

Energy consumption and service quality are not the only possible objectives for optimization of public transport. Road traffic, for example, is a source of noise and local air pollution and occupies significant shares of urban space. Bi-modal ride-pooling reduces traffic by use of shared shuttles, and by directing certain trips towards trains. We quantify this reduction by introducing the relative bi-modal traffic $\traffic$ as the ratio of the number of on-road vehicles necessary for bi-modal transportation (i.e., shuttles) to the number of MIV (i.e., cars) needed to serve the same demand. We have (see \SuppInf~for details)
\begin{equation}\label{eq:traffic_bi}
    \traffic = \eta^{-1}{(1+F)\, \tilde D_\mathrm{shuttle}}\,.
\end{equation}
\begin{figure}[ht]
\centering
\includegraphics{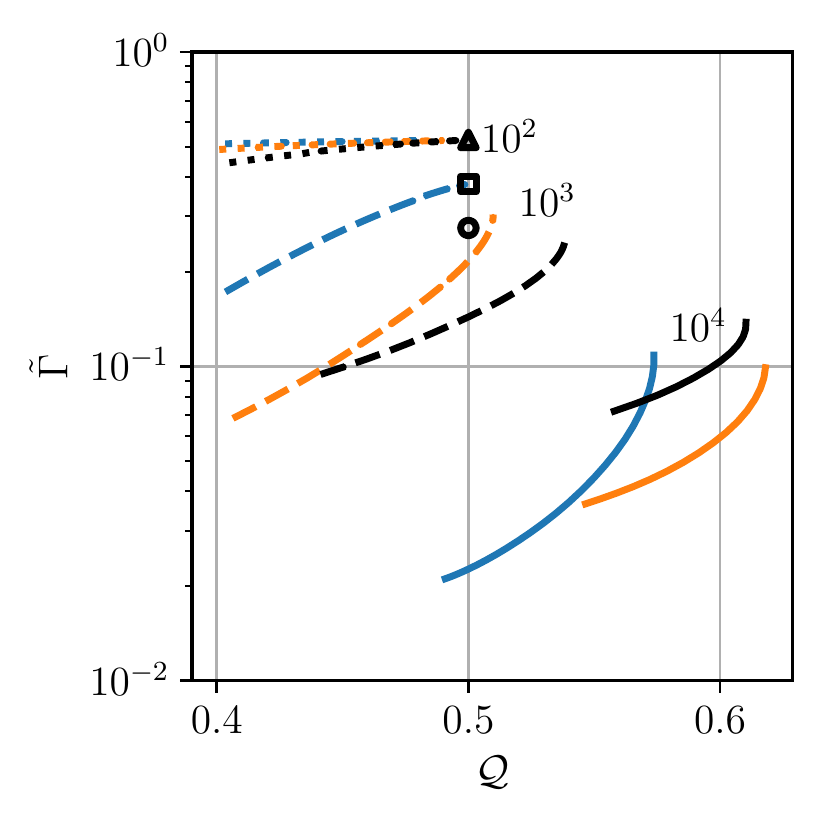}
        \caption{\textbf{Traffic volume in bi-modal transit.}~
        Relative bi-modal traffic $\tilde{\Gamma}$, as defined in Eq.~\ref{eq:traffic_bi}, determined along the Pareto fronts in Fig.~\ref{fig:pareto-intro-miv}c, against corresponding service quality $\quality$. Data are presented for $\Lambda = 10^2$ (triangle, dotted), $\Lambda = 10^3$ (square, dashed), and $\Lambda = 10^4$ (circle, solid). Symbols represent uni-modal traffic volume, $1/\eta$. Color code is as in Fig.~\ref{fig:pareto-intro-miv}, i.e., $\tilde{\ell} = \{0.2, 0.4, 0.8\}$ in blue, orange, and black, respectively.}
        \label{fig:mean and std of nets}
\label{fig:relative_traffic}
\end{figure}
In Fig.~\ref{fig:relative_traffic}, $\tilde{\Gamma}$ is plotted against service quality $\quality$ as determined along the degenerate Pareto fronts shown in Fig.~\ref{fig:pareto-intro-miv}c. For low demand $(\Lambda = 10^2)$, uni-modal (shuttles only) transportation allows for about $50\%$ reduction in traffic as compared to MIV (red triangle). Note the logarithmic scale, which is necessary to display all data properly. Bi-modal transportation allows for further reduction in traffic at the cost of service quality. For intermediate and high demand $(\Lambda = \{10^3,10^4\})$, uni-modal transportation allows for about $70\%$ to $80\%$ reduction in traffic as compared to MIV (square, circle). In these cases, bi-modal transportation allows for truly dramatic reductions in traffic ($>95\%$), at equal or even higher service quality than for uni-modal transport. Combining this finding with typical parameter values in Tab.~\ref{tab:demands}, we recognize that even in rural settings, traffic volume is expected to decrease by an order of magnitude, relative to MIV. In urban environment, traffic volume may even be reduced by more than a factor of ten.

\section{Discussion}
 
Our results indicate that bi-modal public transportation systems have the potential to provide on-demand door-to-door service with a quality (i.e., transit times) superior to customary public transportation, while at the same time consuming only a fraction of the energy a corresponding fleet of MIV would require, and with a road traffic volume reduced by orders of magnitude. We have demonstrated that the marriage of shuttles with trains can improve the shuttle efficiency and dramatically reduce traffic volume. 
 
It is important to note that our results represent only a lower bound on bi-modal performance, in particular as far as service quality is concerned, for several reasons. First, we have based the decision on the type of transport service (uni-modal or bi-modal) on a single scalar parameter, $\dcut$, which amounts to representing the decision process by a binary-valued function of a single scalar variable, $d\longrightarrow\{0,1\}$.
The true structure would be a binary-valued field $\Phi$ on the four-dimensional space of the pick-up and drop-off coordinates, $\Phi\colon(x_p,y_p,x_d,y_d)\longrightarrow\{0,1\}$.
This would be extremely cumbersome to study in a statistical manner.
However, in a real system, data on $\Phi(x_p,y_p,x_d,y_d)$ are being collected on a daily basis, such that over time the system can be ever improving its performance over the data we have presented here.

Second, bi-modal service of the kind studied here potentially provides door-to-door service, while MIV involves the search for parking space, which was completely disregarded in our study due to lack of reliable data. This can be quite significant, and fully adds to the MIV transit time, thus further improving on the relative service quality, $\mathcal{Q}$, of bi-modal service.

Moreover, it should be mentioned that riding the bi-modal transport system neither involves having to drive nor taking care of vehicle maintenance. In summary, it appears that bi-modal public transport systems have  the potential of outperforming the MIV in  a number of respects.

Future work should aim to assess the robustness and scalability of bi-modal transportation on real street networks and diverse network topologies, while adopting more sophisticated dispatching strategies. One aspect that yet remains to be explored is the user adoption of the bi-modal transportation based on incentives, customer convenience, and customer preferences. Recent results from experimental social science, however,  point into a very favorable direction \cite{Avermann2019,Nyga2020,Soerensen2021}.

\section*{Acknowledgement}
The authors would like to appreciate fruitful discussions with  Tariq Baig-Meininghaus, Michael Patscheke and Prakhar Godara.

\bibliography{Bimodal_main}

\begin{thebibliography}{66}%
\makeatletter
\providecommand \@ifxundefined [1]{%
 \@ifx{#1\undefined}
}%
\providecommand \@ifnum [1]{%
 \ifnum #1\expandafter \@firstoftwo
 \else \expandafter \@secondoftwo
 \fi
}%
\providecommand \@ifx [1]{%
 \ifx #1\expandafter \@firstoftwo
 \else \expandafter \@secondoftwo
 \fi
}%
\providecommand \natexlab [1]{#1}%
\providecommand \enquote  [1]{``#1''}%
\providecommand \bibnamefont  [1]{#1}%
\providecommand \bibfnamefont [1]{#1}%
\providecommand \citenamefont [1]{#1}%
\providecommand \href@noop [0]{\@secondoftwo}%
\providecommand \href [0]{\begingroup \@sanitize@url \@href}%
\providecommand \@href[1]{\@@startlink{#1}\@@href}%
\providecommand \@@href[1]{\endgroup#1\@@endlink}%
\providecommand \@sanitize@url [0]{\catcode `\\12\catcode `\$12\catcode
  `\&12\catcode `\#12\catcode `\^12\catcode `\_12\catcode `\%12\relax}%
\providecommand \@@startlink[1]{}%
\providecommand \@@endlink[0]{}%
\providecommand \url  [0]{\begingroup\@sanitize@url \@url }%
\providecommand \@url [1]{\endgroup\@href {#1}{\urlprefix }}%
\providecommand \urlprefix  [0]{URL }%
\providecommand \Eprint [0]{\href }%
\providecommand \doibase [0]{http://dx.doi.org/}%
\providecommand \selectlanguage [0]{\@gobble}%
\providecommand \bibinfo  [0]{\@secondoftwo}%
\providecommand \bibfield  [0]{\@secondoftwo}%
\providecommand \translation [1]{[#1]}%
\providecommand \BibitemOpen [0]{}%
\providecommand \bibitemStop [0]{}%
\providecommand \bibitemNoStop [0]{.\EOS\space}%
\providecommand \EOS [0]{\spacefactor3000\relax}%
\providecommand \BibitemShut  [1]{\csname bibitem#1\endcsname}%
\let\auto@bib@innerbib\@empty
\bibitem [{\citenamefont {Fan}\ \emph {et~al.}(2018)\citenamefont {Fan},
  \citenamefont {Perry}, \citenamefont {Klemeš},\ and\ \citenamefont
  {Lee}}]{air_emission_review}%
  \BibitemOpen
  \bibfield  {author} {\bibinfo {author} {\bibfnamefont {Y.~V.}\ \bibnamefont
  {Fan}}, \bibinfo {author} {\bibfnamefont {S.}~\bibnamefont {Perry}}, \bibinfo
  {author} {\bibfnamefont {J.~J.}\ \bibnamefont {Klemeš}}, \ and\ \bibinfo
  {author} {\bibfnamefont {C.~T.}\ \bibnamefont {Lee}},\ }\href {\doibase
  https://doi.org/10.1016/j.jclepro.2018.05.151} {\bibfield  {journal}
  {\bibinfo  {journal} {Journal of Cleaner Production}\ }\textbf {\bibinfo
  {volume} {194}},\ \bibinfo {pages} {673} (\bibinfo {year}
  {2018})}\BibitemShut {NoStop}%
\bibitem [{\citenamefont {Kontovas}\ and\ \citenamefont
  {Psaraftis}(2016)}]{air_emissions_Kontovas2016}%
  \BibitemOpen
  \bibfield  {author} {\bibinfo {author} {\bibfnamefont {C.~A.}\ \bibnamefont
  {Kontovas}}\ and\ \bibinfo {author} {\bibfnamefont {H.~N.}\ \bibnamefont
  {Psaraftis}},\ }\enquote {\bibinfo {title} {Transportation emissions: Some
  basics},}\ in\ \href {\doibase 10.1007/978-3-319-17175-3_2} {\emph {\bibinfo
  {booktitle} {Green Transportation Logistics: The Quest for Win-Win
  Solutions}}},\ \bibinfo {editor} {edited by\ \bibinfo {editor} {\bibfnamefont
  {H.~N.}\ \bibnamefont {Psaraftis}}}\ (\bibinfo  {publisher} {Springer
  International Publishing},\ \bibinfo {address} {Cham},\ \bibinfo {year}
  {2016})\BibitemShut {NoStop}%
\bibitem [{\citenamefont {Douglas}\ \emph {et~al.}(2011)\citenamefont
  {Douglas}, \citenamefont {Watkins}, \citenamefont {Gorman},\ and\
  \citenamefont {Higgins}}]{reliance_MIV_Douglas}%
  \BibitemOpen
  \bibfield  {author} {\bibinfo {author} {\bibfnamefont {M.~J.}\ \bibnamefont
  {Douglas}}, \bibinfo {author} {\bibfnamefont {S.~J.}\ \bibnamefont
  {Watkins}}, \bibinfo {author} {\bibfnamefont {D.~R.}\ \bibnamefont {Gorman}},
  \ and\ \bibinfo {author} {\bibfnamefont {M.}~\bibnamefont {Higgins}},\ }\href
  {\doibase 10.1093/pubmed/fdr032} {\bibfield  {journal} {\bibinfo  {journal}
  {Journal of Public Health}\ }\textbf {\bibinfo {volume} {33}},\ \bibinfo
  {pages} {160} (\bibinfo {year} {2011})},\ \Eprint
  {http://arxiv.org/abs/https://academic.oup.com/jpubhealth/article-pdf/33/2/160/4513901/fdr032.pdf}
  {https://academic.oup.com/jpubhealth/article-pdf/33/2/160/4513901/fdr032.pdf}
  \BibitemShut {NoStop}%
\bibitem [{\citenamefont {Newman}\ and\ \citenamefont
  {Kenworthy}(1989)}]{reliance_MIV_newman1989cities}%
  \BibitemOpen
  \bibfield  {author} {\bibinfo {author} {\bibfnamefont {P.~W.~G.}\
  \bibnamefont {Newman}}\ and\ \bibinfo {author} {\bibfnamefont {J.~R.}\
  \bibnamefont {Kenworthy}},\ }\href@noop {} {\emph {\bibinfo {title} {Cities
  and automobile dependence : a sourcebook}}}\ (\bibinfo  {publisher} {Gower
  Technical Aldershot, Hants., England ; Brookfield, Vt., USA},\ \bibinfo
  {year} {1989})\BibitemShut {NoStop}%
\bibitem [{\citenamefont {Mingardo}\ \emph {et~al.}(2022)\citenamefont
  {Mingardo}, \citenamefont {Vermeulen},\ and\ \citenamefont
  {Bornioli}}]{parking_giuliano}%
  \BibitemOpen
  \bibfield  {author} {\bibinfo {author} {\bibfnamefont {G.}~\bibnamefont
  {Mingardo}}, \bibinfo {author} {\bibfnamefont {S.}~\bibnamefont {Vermeulen}},
  \ and\ \bibinfo {author} {\bibfnamefont {A.}~\bibnamefont {Bornioli}},\
  }\href {\doibase https://doi.org/10.1016/j.tra.2022.01.005} {\bibfield
  {journal} {\bibinfo  {journal} {Transportation Research Part A: Policy and
  Practice}\ }\textbf {\bibinfo {volume} {157}},\ \bibinfo {pages} {185}
  (\bibinfo {year} {2022})}\BibitemShut {NoStop}%
\bibitem [{\citenamefont {Manville}\ and\ \citenamefont
  {Shoup}(2005)}]{manville2005parking}%
  \BibitemOpen
  \bibfield  {author} {\bibinfo {author} {\bibfnamefont {M.}~\bibnamefont
  {Manville}}\ and\ \bibinfo {author} {\bibfnamefont {D.}~\bibnamefont
  {Shoup}},\ }\href@noop {} {\bibfield  {journal} {\bibinfo  {journal} {Journal
  of urban planning and development}\ }\textbf {\bibinfo {volume} {131}},\
  \bibinfo {pages} {233} (\bibinfo {year} {2005})}\BibitemShut {NoStop}%
\bibitem [{\citenamefont {Park}\ \emph {et~al.}(2012)\citenamefont {Park},
  \citenamefont {Kim}, \citenamefont {Park},\ and\ \citenamefont
  {Kim}}]{logistic_dongjoo}%
  \BibitemOpen
  \bibfield  {author} {\bibinfo {author} {\bibfnamefont {D.}~\bibnamefont
  {Park}}, \bibinfo {author} {\bibfnamefont {N.~S.}\ \bibnamefont {Kim}},
  \bibinfo {author} {\bibfnamefont {H.}~\bibnamefont {Park}}, \ and\ \bibinfo
  {author} {\bibfnamefont {K.}~\bibnamefont {Kim}},\ }\href {\doibase
  10.1080/12265934.2012.668322} {\bibfield  {journal} {\bibinfo  {journal}
  {International Journal of Urban Sciences}\ }\textbf {\bibinfo {volume}
  {16}},\ \bibinfo {pages} {85} (\bibinfo {year} {2012})},\ \Eprint
  {http://arxiv.org/abs/https://doi.org/10.1080/12265934.2012.668322}
  {https://doi.org/10.1080/12265934.2012.668322} \BibitemShut {NoStop}%
\bibitem [{\citenamefont {Jang}\ \emph {et~al.}(2016)\citenamefont {Jang},
  \citenamefont {Jeong},\ and\ \citenamefont {Lee}}]{logistic_Jang}%
  \BibitemOpen
  \bibfield  {author} {\bibinfo {author} {\bibfnamefont {Y.~J.}\ \bibnamefont
  {Jang}}, \bibinfo {author} {\bibfnamefont {S.}~\bibnamefont {Jeong}}, \ and\
  \bibinfo {author} {\bibfnamefont {M.~S.}\ \bibnamefont {Lee}},\ }\href
  {\doibase 10.3390/en9070483} {\bibfield  {journal} {\bibinfo  {journal}
  {Energies}\ }\textbf {\bibinfo {volume} {9}} (\bibinfo {year} {2016}),\
  10.3390/en9070483}\BibitemShut {NoStop}%
\bibitem [{\citenamefont {Swenseth}\ and\ \citenamefont
  {Godfrey}(2002)}]{logistic_scott}%
  \BibitemOpen
  \bibfield  {author} {\bibinfo {author} {\bibfnamefont {S.~R.}\ \bibnamefont
  {Swenseth}}\ and\ \bibinfo {author} {\bibfnamefont {M.~R.}\ \bibnamefont
  {Godfrey}},\ }\href {\doibase https://doi.org/10.1016/S0925-5273(01)00230-4}
  {\bibfield  {journal} {\bibinfo  {journal} {International Journal of
  Production Economics}\ }\textbf {\bibinfo {volume} {77}},\ \bibinfo {pages}
  {113} (\bibinfo {year} {2002})}\BibitemShut {NoStop}%
\bibitem [{eea(2020)}]{eeaoccupancy}%
  \BibitemOpen
  \href@noop {} {\enquote {\bibinfo {title} {European environment agency: Are
  we moving in the right direction? indicators on transport and environmental
  integration in the eu},}\ }\bibinfo {howpublished}
  {\url{https://www.eea.europa.eu/ds_resolveuid/0c1c4a6acf289ffdefa1876ea5d60f07}}
  (\bibinfo {year} {2020}),\ \bibinfo {note} {accessed: 2022-07-14}\BibitemShut
  {NoStop}%
\bibitem [{\citenamefont {Joireman}\ \emph {et~al.}(2004)\citenamefont
  {Joireman}, \citenamefont {Van~Lange},\ and\ \citenamefont
  {Van~Vugt}}]{environment_joireman2004cares}%
  \BibitemOpen
  \bibfield  {author} {\bibinfo {author} {\bibfnamefont {J.~A.}\ \bibnamefont
  {Joireman}}, \bibinfo {author} {\bibfnamefont {P.~A.}\ \bibnamefont
  {Van~Lange}}, \ and\ \bibinfo {author} {\bibfnamefont {M.}~\bibnamefont
  {Van~Vugt}},\ }\href@noop {} {\bibfield  {journal} {\bibinfo  {journal}
  {Environment and behavior}\ }\textbf {\bibinfo {volume} {36}},\ \bibinfo
  {pages} {187} (\bibinfo {year} {2004})}\BibitemShut {NoStop}%
\bibitem [{\citenamefont {Chin}(1996)}]{traffic_singapore}%
  \BibitemOpen
  \bibfield  {author} {\bibinfo {author} {\bibfnamefont {A.~T.}\ \bibnamefont
  {Chin}},\ }\href {\doibase https://doi.org/10.1016/1352-2310(95)00173-5}
  {\bibfield  {journal} {\bibinfo  {journal} {Atmospheric Environment}\
  }\textbf {\bibinfo {volume} {30}},\ \bibinfo {pages} {787} (\bibinfo {year}
  {1996})},\ \bibinfo {note} {supercities: Environment Quality and Sustainable
  Development}\BibitemShut {NoStop}%
\bibitem [{\citenamefont {Arnott}\ and\ \citenamefont {Small}(1994)}]{traffic}%
  \BibitemOpen
  \bibfield  {author} {\bibinfo {author} {\bibfnamefont {R.}~\bibnamefont
  {Arnott}}\ and\ \bibinfo {author} {\bibfnamefont {K.}~\bibnamefont {Small}},\
  }\href {http://www.jstor.org/stable/29775281} {\bibfield  {journal} {\bibinfo
   {journal} {American Scientist}\ }\textbf {\bibinfo {volume} {82}},\ \bibinfo
  {pages} {446} (\bibinfo {year} {1994})}\BibitemShut {NoStop}%
\bibitem [{\citenamefont {{Ko\'zlak, Aleksandra}}\ and\ \citenamefont {{Wach,
  Dagmara}}(2018)}]{traffic_Poland}%
  \BibitemOpen
  \bibfield  {author} {\bibinfo {author} {\bibnamefont {{Ko\'zlak,
  Aleksandra}}}\ and\ \bibinfo {author} {\bibnamefont {{Wach, Dagmara}}},\
  }\href {\doibase 10.1051/shsconf/20185701019} {\bibfield  {journal} {\bibinfo
   {journal} {SHS Web Conf.}\ }\textbf {\bibinfo {volume} {57}},\ \bibinfo
  {pages} {01019} (\bibinfo {year} {2018})}\BibitemShut {NoStop}%
\bibitem [{\citenamefont {Barth}\ and\ \citenamefont
  {Boriboonsomsin}(2009)}]{traffic_greenhouse}%
  \BibitemOpen
  \bibfield  {author} {\bibinfo {author} {\bibfnamefont {M.}~\bibnamefont
  {Barth}}\ and\ \bibinfo {author} {\bibfnamefont {K.}~\bibnamefont
  {Boriboonsomsin}},\ }\href@noop {} {\bibfield  {journal} {\bibinfo  {journal}
  {Access Magazine}\ }\textbf {\bibinfo {volume} {1}},\ \bibinfo {pages} {2}
  (\bibinfo {year} {2009})}\BibitemShut {NoStop}%
\bibitem [{\citenamefont {Agency}(2020)}]{eeaair}%
  \BibitemOpen
  \bibfield  {author} {\bibinfo {author} {\bibfnamefont {E.~E.}\ \bibnamefont
  {Agency}},\ }\href@noop {} {\enquote {\bibinfo {title} {Air quality in europe
  — 2020 report},}\ }\bibinfo {howpublished}
  {\url{https://www.eea.europa.eu/publications/air-quality-in-europe-2020-report}}
  (\bibinfo {year} {2020}),\ \bibinfo {note} {accessed: 2022-05-18}\BibitemShut
  {NoStop}%
\bibitem [{\citenamefont {Caiazzo}\ \emph {et~al.}(2013)\citenamefont
  {Caiazzo}, \citenamefont {Ashok}, \citenamefont {Waitz}, \citenamefont
  {Yim},\ and\ \citenamefont {Barrett}}]{ashok_pollution}%
  \BibitemOpen
  \bibfield  {author} {\bibinfo {author} {\bibfnamefont {F.}~\bibnamefont
  {Caiazzo}}, \bibinfo {author} {\bibfnamefont {A.}~\bibnamefont {Ashok}},
  \bibinfo {author} {\bibfnamefont {I.~A.}\ \bibnamefont {Waitz}}, \bibinfo
  {author} {\bibfnamefont {S.~H.}\ \bibnamefont {Yim}}, \ and\ \bibinfo
  {author} {\bibfnamefont {S.~R.}\ \bibnamefont {Barrett}},\ }\href {\doibase
  https://doi.org/10.1016/j.atmosenv.2013.05.081} {\bibfield  {journal}
  {\bibinfo  {journal} {Atmospheric Environment}\ }\textbf {\bibinfo {volume}
  {79}},\ \bibinfo {pages} {198} (\bibinfo {year} {2013})}\BibitemShut
  {NoStop}%
\bibitem [{\citenamefont {MacKenzie}\ \emph {et~al.}(2014)\citenamefont
  {MacKenzie}, \citenamefont {Zoepf},\ and\ \citenamefont
  {Heywood}}]{mackenzie2014determinants}%
  \BibitemOpen
  \bibfield  {author} {\bibinfo {author} {\bibfnamefont {D.}~\bibnamefont
  {MacKenzie}}, \bibinfo {author} {\bibfnamefont {S.}~\bibnamefont {Zoepf}}, \
  and\ \bibinfo {author} {\bibfnamefont {J.}~\bibnamefont {Heywood}},\
  }\href@noop {} {\bibfield  {journal} {\bibinfo  {journal} {Int. J. Veh. Des}\
  }\textbf {\bibinfo {volume} {65}},\ \bibinfo {pages} {73} (\bibinfo {year}
  {2014})}\BibitemShut {NoStop}%
\bibitem [{\citenamefont {Tachet}\ \emph {et~al.}(2017)\citenamefont {Tachet},
  \citenamefont {Sagarra}, \citenamefont {Santi}, \citenamefont {Resta},
  \citenamefont {Szell}, \citenamefont {Strogatz},\ and\ \citenamefont
  {Ratti}}]{tachet_scaling_2017}%
  \BibitemOpen
  \bibfield  {author} {\bibinfo {author} {\bibfnamefont {R.}~\bibnamefont
  {Tachet}}, \bibinfo {author} {\bibfnamefont {O.}~\bibnamefont {Sagarra}},
  \bibinfo {author} {\bibfnamefont {P.}~\bibnamefont {Santi}}, \bibinfo
  {author} {\bibfnamefont {G.}~\bibnamefont {Resta}}, \bibinfo {author}
  {\bibfnamefont {M.}~\bibnamefont {Szell}}, \bibinfo {author} {\bibfnamefont
  {S.~H.}\ \bibnamefont {Strogatz}}, \ and\ \bibinfo {author} {\bibfnamefont
  {C.}~\bibnamefont {Ratti}},\ }\href@noop {} {\bibfield  {journal} {\bibinfo
  {journal} {Scientific Reports}\ }\textbf {\bibinfo {volume} {7}},\ \bibinfo
  {pages} {1} (\bibinfo {year} {2017})}\BibitemShut {NoStop}%
\bibitem [{MIV(2022)}]{MIV_dominance}%
  \BibitemOpen
  \href@noop {} {\enquote {\bibinfo {title} {Passenger mobility statistics},}\
  }\bibinfo {howpublished}
  {\url{https://ec.europa.eu/eurostat/statistics-explained/index.php?title=Passenger_mobility_statistics#Urban_trips}}
  (\bibinfo {year} {2022}),\ \bibinfo {note} {accessed: 2022-09-29}\BibitemShut
  {NoStop}%
\bibitem [{\citenamefont {Fiorello}\ \emph {et~al.}(2016)\citenamefont
  {Fiorello}, \citenamefont {Martino}, \citenamefont {Zani}, \citenamefont
  {Christidis},\ and\ \citenamefont {Navajas-Cawood}}]{MIV_dominance_2}%
  \BibitemOpen
  \bibfield  {author} {\bibinfo {author} {\bibfnamefont {D.}~\bibnamefont
  {Fiorello}}, \bibinfo {author} {\bibfnamefont {A.}~\bibnamefont {Martino}},
  \bibinfo {author} {\bibfnamefont {L.}~\bibnamefont {Zani}}, \bibinfo {author}
  {\bibfnamefont {P.}~\bibnamefont {Christidis}}, \ and\ \bibinfo {author}
  {\bibfnamefont {E.}~\bibnamefont {Navajas-Cawood}},\ }\href@noop {}
  {\bibfield  {journal} {\bibinfo  {journal} {Transportation research
  procedia}\ }\textbf {\bibinfo {volume} {14}},\ \bibinfo {pages} {1104}
  (\bibinfo {year} {2016})}\BibitemShut {NoStop}%
\bibitem [{\citenamefont {Kent}(2013)}]{kent2013secured}%
  \BibitemOpen
  \bibfield  {author} {\bibinfo {author} {\bibfnamefont {J.}~\bibnamefont
  {Kent}},\ }\emph {\bibinfo {title} {Secured by automobility: why does the
  private car continue to dominate transport practices?}},\ \href@noop {}
  {Ph.D. thesis},\ \bibinfo  {school} {UNSW Sydney} (\bibinfo {year}
  {2013})\BibitemShut {NoStop}%
\bibitem [{\citenamefont {Shaheen}\ and\ \citenamefont
  {Cohen}(2019)}]{pooling_north_america}%
  \BibitemOpen
  \bibfield  {author} {\bibinfo {author} {\bibfnamefont {S.}~\bibnamefont
  {Shaheen}}\ and\ \bibinfo {author} {\bibfnamefont {A.}~\bibnamefont
  {Cohen}},\ }\href {\doibase 10.1080/01441647.2018.1497728} {\bibfield
  {journal} {\bibinfo  {journal} {Transport Reviews}\ }\textbf {\bibinfo
  {volume} {39}},\ \bibinfo {pages} {427} (\bibinfo {year} {2019})},\ \Eprint
  {http://arxiv.org/abs/https://doi.org/10.1080/01441647.2018.1497728}
  {https://doi.org/10.1080/01441647.2018.1497728} \BibitemShut {NoStop}%
\bibitem [{\citenamefont {Zwick}\ \emph {et~al.}(2021)\citenamefont {Zwick},
  \citenamefont {Kuehnel}, \citenamefont {Moeckel},\ and\ \citenamefont
  {Axhausen}}]{munich}%
  \BibitemOpen
  \bibfield  {author} {\bibinfo {author} {\bibfnamefont {F.}~\bibnamefont
  {Zwick}}, \bibinfo {author} {\bibfnamefont {N.}~\bibnamefont {Kuehnel}},
  \bibinfo {author} {\bibfnamefont {R.}~\bibnamefont {Moeckel}}, \ and\
  \bibinfo {author} {\bibfnamefont {K.~W.}\ \bibnamefont {Axhausen}},\ }\href
  {\doibase https://doi.org/10.1016/j.procs.2021.03.083} {\bibfield  {journal}
  {\bibinfo  {journal} {Procedia Computer Science}\ }\textbf {\bibinfo {volume}
  {184}},\ \bibinfo {pages} {662} (\bibinfo {year} {2021})},\ \bibinfo {note}
  {the 12th International Conference on Ambient Systems, Networks and
  Technologies (ANT) / The 4th International Conference on Emerging Data and
  Industry 4.0 (EDI40) / Affiliated Workshops}\BibitemShut {NoStop}%
\bibitem [{\citenamefont {Chen}\ \emph {et~al.}(2017)\citenamefont {Chen},
  \citenamefont {Jauhri},\ and\ \citenamefont {Shen}}]{data_driven_pooling}%
  \BibitemOpen
  \bibfield  {author} {\bibinfo {author} {\bibfnamefont {M.~H.}\ \bibnamefont
  {Chen}}, \bibinfo {author} {\bibfnamefont {A.}~\bibnamefont {Jauhri}}, \ and\
  \bibinfo {author} {\bibfnamefont {J.~P.}\ \bibnamefont {Shen}},\ }in\ \href
  {\doibase 10.1145/3151547.3151549} {\emph {\bibinfo {booktitle} {Proceedings
  of the 10th ACM SIGSPATIAL Workshop on Computational Transportation
  Science}}},\ \bibinfo {series and number} {IWCTS'17}\ (\bibinfo  {publisher}
  {Association for Computing Machinery},\ \bibinfo {address} {New York, NY,
  USA},\ \bibinfo {year} {2017})\ p.\ \bibinfo {pages} {7–12}\BibitemShut
  {NoStop}%
\bibitem [{\citenamefont {Merlin}(2019)}]{merlin2019transportation}%
  \BibitemOpen
  \bibfield  {author} {\bibinfo {author} {\bibfnamefont {L.~A.}\ \bibnamefont
  {Merlin}},\ }\href@noop {} {\bibfield  {journal} {\bibinfo  {journal}
  {Journal of the American Planning Association}\ }\textbf {\bibinfo {volume}
  {85}},\ \bibinfo {pages} {501} (\bibinfo {year} {2019})}\BibitemShut
  {NoStop}%
\bibitem [{\citenamefont {Pietrzak}(2016)}]{pitrzak}%
  \BibitemOpen
  \bibfield  {author} {\bibinfo {author} {\bibfnamefont {K.}~\bibnamefont
  {Pietrzak}},\ }\href@noop {} {\bibfield  {journal} {\bibinfo  {journal}
  {Transportation Research Procedia}\ }\textbf {\bibinfo {volume} {16}},\
  \bibinfo {pages} {464–472} (\bibinfo {year} {2016})}\BibitemShut {NoStop}%
\bibitem [{\citenamefont {Pietrzak}(2019)}]{pitrzak_cities}%
  \BibitemOpen
  \bibfield  {author} {\bibinfo {author} {\bibfnamefont {K.}~\bibnamefont
  {Pietrzak}, \bibfnamefont {O.~\&~Pietrzak}},\ }\href@noop {} {\bibfield
  {journal} {\bibinfo  {journal} {Transportation Research Procedia}\ }\textbf
  {\bibinfo {volume} {39}},\ \bibinfo {pages} {405–416} (\bibinfo {year}
  {2019})}\BibitemShut {NoStop}%
\bibitem [{\citenamefont {Ferbrache}(2017)}]{ferbrache_cities}%
  \BibitemOpen
  \bibfield  {author} {\bibinfo {author} {\bibfnamefont {R.~D.}\ \bibnamefont
  {Ferbrache}, \bibfnamefont {F.~\&~Knowles}},\ }\href@noop {} {\bibfield
  {journal} {\bibinfo  {journal} {Geoforum}\ }\textbf {\bibinfo {volume}
  {80}},\ \bibinfo {pages} {103–113} (\bibinfo {year} {2017})}\BibitemShut
  {NoStop}%
\bibitem [{\citenamefont {Kato}\ and\ \citenamefont {Kaneko Y.
  \&~Soyama}(2014)}]{kato}%
  \BibitemOpen
  \bibfield  {author} {\bibinfo {author} {\bibfnamefont {H.}~\bibnamefont
  {Kato}}\ and\ \bibinfo {author} {\bibfnamefont {Y.}~\bibnamefont {Kaneko Y.
  \&~Soyama}},\ }\href@noop {} {\bibfield  {journal} {\bibinfo  {journal}
  {Transport Policy}\ }\textbf {\bibinfo {volume} {35}},\ \bibinfo {pages}
  {202–210} (\bibinfo {year} {2014})}\BibitemShut {NoStop}%
\bibitem [{\citenamefont {Alam}\ \emph {et~al.}(2015)\citenamefont {Alam},
  \citenamefont {Nixon},\ and\ \citenamefont
  {Zhang}}]{Alam2015InvestigatingTD}%
  \BibitemOpen
  \bibfield  {author} {\bibinfo {author} {\bibfnamefont {B.~M.}\ \bibnamefont
  {Alam}}, \bibinfo {author} {\bibfnamefont {H.}~\bibnamefont {Nixon}}, \ and\
  \bibinfo {author} {\bibfnamefont {Q.}~\bibnamefont {Zhang}}\ }(\bibinfo
  {year} {2015})\BibitemShut {NoStop}%
\bibitem [{\citenamefont {Herminghaus}(2019)}]{herminghaus_mean_2019}%
  \BibitemOpen
  \bibfield  {author} {\bibinfo {author} {\bibfnamefont {S.}~\bibnamefont
  {Herminghaus}},\ }\href@noop {} {\bibfield  {journal} {\bibinfo  {journal}
  {Transportation Research Part A: Policy and Practice}\ }\textbf {\bibinfo
  {volume} {119}},\ \bibinfo {pages} {15} (\bibinfo {year} {2019})}\BibitemShut
  {NoStop}%
\bibitem [{\citenamefont {Alonso-Mora}\ \emph {et~al.}(2017)\citenamefont
  {Alonso-Mora}, \citenamefont {Samaranayake}, \citenamefont {Wallar},
  \citenamefont {Frazzoli},\ and\ \citenamefont
  {Rus}}]{alonso-mora_-demand_2017}%
  \BibitemOpen
  \bibfield  {author} {\bibinfo {author} {\bibfnamefont {J.}~\bibnamefont
  {Alonso-Mora}}, \bibinfo {author} {\bibfnamefont {S.}~\bibnamefont
  {Samaranayake}}, \bibinfo {author} {\bibfnamefont {A.}~\bibnamefont
  {Wallar}}, \bibinfo {author} {\bibfnamefont {E.}~\bibnamefont {Frazzoli}}, \
  and\ \bibinfo {author} {\bibfnamefont {D.}~\bibnamefont {Rus}},\ }\href@noop
  {} {\bibfield  {journal} {\bibinfo  {journal} {Proceedings of the National
  Academy of Sciences}\ }\textbf {\bibinfo {volume} {114}},\ \bibinfo {pages}
  {462} (\bibinfo {year} {2017})}\BibitemShut {NoStop}%
\bibitem [{\citenamefont {Lobel}\ and\ \citenamefont
  {Martin}(2020)}]{lobel_detours_2020}%
  \BibitemOpen
  \bibfield  {author} {\bibinfo {author} {\bibfnamefont {I.}~\bibnamefont
  {Lobel}}\ and\ \bibinfo {author} {\bibfnamefont {S.}~\bibnamefont {Martin}},\
  }\href@noop {} {\bibfield  {journal} {\bibinfo  {journal} {Available at SSRN
  3711072}\ } (\bibinfo {year} {2020})}\BibitemShut {NoStop}%
\bibitem [{\citenamefont {Daganzo}\ \emph {et~al.}(2020)\citenamefont
  {Daganzo}, \citenamefont {Ouyang},\ and\ \citenamefont
  {Yang}}]{daganzo2020analysis}%
  \BibitemOpen
  \bibfield  {author} {\bibinfo {author} {\bibfnamefont {C.~F.}\ \bibnamefont
  {Daganzo}}, \bibinfo {author} {\bibfnamefont {Y.}~\bibnamefont {Ouyang}}, \
  and\ \bibinfo {author} {\bibfnamefont {H.}~\bibnamefont {Yang}},\ }\href@noop
  {} {\bibfield  {journal} {\bibinfo  {journal} {Transportation Research Part
  B: Methodological}\ }\textbf {\bibinfo {volume} {140}},\ \bibinfo {pages}
  {130} (\bibinfo {year} {2020})}\BibitemShut {NoStop}%
\bibitem [{\citenamefont {Molkenthin}\ \emph {et~al.}(2020)\citenamefont
  {Molkenthin}, \citenamefont {Schr\"oder},\ and\ \citenamefont
  {Timme}}]{nora_scaling_2020}%
  \BibitemOpen
  \bibfield  {author} {\bibinfo {author} {\bibfnamefont {N.}~\bibnamefont
  {Molkenthin}}, \bibinfo {author} {\bibfnamefont {M.}~\bibnamefont
  {Schr\"oder}}, \ and\ \bibinfo {author} {\bibfnamefont {M.}~\bibnamefont
  {Timme}},\ }\href {\doibase 10.1103/PhysRevLett.125.248302} {\bibfield
  {journal} {\bibinfo  {journal} {Phys. Rev. Lett.}\ }\textbf {\bibinfo
  {volume} {125}},\ \bibinfo {pages} {248302} (\bibinfo {year}
  {2020})}\BibitemShut {NoStop}%
\bibitem [{\citenamefont {Salonen}(2013)}]{salonen}%
  \BibitemOpen
  \bibfield  {author} {\bibinfo {author} {\bibfnamefont {T.}~\bibnamefont
  {Salonen}, \bibfnamefont {M.~\&~Toivonen}},\ }\href@noop {} {\bibfield
  {journal} {\bibinfo  {journal} {Journal of Transport Geography}\ }\textbf
  {\bibinfo {volume} {31}},\ \bibinfo {pages} {143–153} (\bibinfo {year}
  {2013})}\BibitemShut {NoStop}%
\bibitem [{\citenamefont {Sorge}\ \emph {et~al.}(2015)\citenamefont {Sorge},
  \citenamefont {Manik}, \citenamefont {Herminghaus},\ and\ \citenamefont
  {Timme}}]{sorge}%
  \BibitemOpen
  \bibfield  {author} {\bibinfo {author} {\bibfnamefont {A.}~\bibnamefont
  {Sorge}}, \bibinfo {author} {\bibfnamefont {D.}~\bibnamefont {Manik}},
  \bibinfo {author} {\bibfnamefont {S.}~\bibnamefont {Herminghaus}}, \ and\
  \bibinfo {author} {\bibfnamefont {M.}~\bibnamefont {Timme}},\ }in\ \href@noop
  {} {\emph {\bibinfo {booktitle} {Proceedings of the 2015 Winter Simulation
  Conference}}},\ \bibinfo {series and number} {WSC '15}\ (\bibinfo
  {publisher} {IEEE Press},\ \bibinfo {year} {2015})\ p.\ \bibinfo {pages}
  {2800–2811}\BibitemShut {NoStop}%
\bibitem [{\citenamefont {Santi}\ \emph {et~al.}(2014)\citenamefont {Santi},
  \citenamefont {Resta}, \citenamefont {Szell}, \citenamefont {Sobolevsky},
  \citenamefont {Strogatz},\ and\ \citenamefont
  {Ratti}}]{santi_shareability_2014}%
  \BibitemOpen
  \bibfield  {author} {\bibinfo {author} {\bibfnamefont {P.}~\bibnamefont
  {Santi}}, \bibinfo {author} {\bibfnamefont {G.}~\bibnamefont {Resta}},
  \bibinfo {author} {\bibfnamefont {M.}~\bibnamefont {Szell}}, \bibinfo
  {author} {\bibfnamefont {S.}~\bibnamefont {Sobolevsky}}, \bibinfo {author}
  {\bibfnamefont {S.~H.}\ \bibnamefont {Strogatz}}, \ and\ \bibinfo {author}
  {\bibfnamefont {C.}~\bibnamefont {Ratti}},\ }\href {\doibase
  10.1073/pnas.1403657111} {\bibfield  {journal} {\bibinfo  {journal}
  {Proceedings of the National Academy of Sciences}\ }\textbf {\bibinfo
  {volume} {111}},\ \bibinfo {pages} {13290} (\bibinfo {year} {2014})},\
  \Eprint
  {http://arxiv.org/abs/https://www.pnas.org/doi/pdf/10.1073/pnas.1403657111}
  {https://www.pnas.org/doi/pdf/10.1073/pnas.1403657111} \BibitemShut {NoStop}%
\bibitem [{\citenamefont {Vazifeh}\ \emph {et~al.}(2018)\citenamefont
  {Vazifeh}, \citenamefont {Santi}, \citenamefont {Resta}, \citenamefont
  {Strogatz},\ and\ \citenamefont {Ratti}}]{vazifeh_addressing_2018}%
  \BibitemOpen
  \bibfield  {author} {\bibinfo {author} {\bibfnamefont {M.~M.}\ \bibnamefont
  {Vazifeh}}, \bibinfo {author} {\bibfnamefont {P.}~\bibnamefont {Santi}},
  \bibinfo {author} {\bibfnamefont {G.}~\bibnamefont {Resta}}, \bibinfo
  {author} {\bibfnamefont {S.~H.}\ \bibnamefont {Strogatz}}, \ and\ \bibinfo
  {author} {\bibfnamefont {C.}~\bibnamefont {Ratti}},\ }\href@noop {}
  {\bibfield  {journal} {\bibinfo  {journal} {Nature}\ }\textbf {\bibinfo
  {volume} {557}},\ \bibinfo {pages} {534} (\bibinfo {year}
  {2018})}\BibitemShut {NoStop}%
\bibitem [{\citenamefont {Sundt}\ \emph {et~al.}(2021)\citenamefont {Sundt},
  \citenamefont {Luo}, \citenamefont {Vincent}, \citenamefont {Shahabi},\ and\
  \citenamefont {Yin}}]{heuristic_pooling_quality}%
  \BibitemOpen
  \bibfield  {author} {\bibinfo {author} {\bibfnamefont {A.}~\bibnamefont
  {Sundt}}, \bibinfo {author} {\bibfnamefont {Q.}~\bibnamefont {Luo}}, \bibinfo
  {author} {\bibfnamefont {J.}~\bibnamefont {Vincent}}, \bibinfo {author}
  {\bibfnamefont {M.}~\bibnamefont {Shahabi}}, \ and\ \bibinfo {author}
  {\bibfnamefont {Y.}~\bibnamefont {Yin}},\ }\href {\doibase
  10.48550/ARXIV.2107.11318} {\enquote {\bibinfo {title} {Heuristics for
  customer-focused ride-pooling assignment},}\ } (\bibinfo {year}
  {2021})\BibitemShut {NoStop}%
\bibitem [{\citenamefont {Storch}\ \emph {et~al.}(2021)\citenamefont {Storch},
  \citenamefont {Timme},\ and\ \citenamefont
  {Schröder}}]{storch_incentive-driven_2021}%
  \BibitemOpen
  \bibfield  {author} {\bibinfo {author} {\bibfnamefont {D.-M.}\ \bibnamefont
  {Storch}}, \bibinfo {author} {\bibfnamefont {M.}~\bibnamefont {Timme}}, \
  and\ \bibinfo {author} {\bibfnamefont {M.}~\bibnamefont {Schröder}},\ }\href
  {\doibase 10.1038/s41467-021-23287-6} {\bibfield  {journal} {\bibinfo
  {journal} {Nature Communications}\ }\textbf {\bibinfo {volume} {12}},\
  \bibinfo {pages} {3003} (\bibinfo {year} {2021})}\BibitemShut {NoStop}%
\bibitem [{\citenamefont {Wolf}\ \emph {et~al.}(2022)\citenamefont {Wolf},
  \citenamefont {Storch}, \citenamefont {Timme},\ and\ \citenamefont
  {Schr\"oder}}]{storch_symmetry}%
  \BibitemOpen
  \bibfield  {author} {\bibinfo {author} {\bibfnamefont {H.}~\bibnamefont
  {Wolf}}, \bibinfo {author} {\bibfnamefont {D.-M.}\ \bibnamefont {Storch}},
  \bibinfo {author} {\bibfnamefont {M.}~\bibnamefont {Timme}}, \ and\ \bibinfo
  {author} {\bibfnamefont {M.}~\bibnamefont {Schr\"oder}},\ }\href {\doibase
  10.1103/PhysRevE.105.044309} {\bibfield  {journal} {\bibinfo  {journal}
  {Phys. Rev. E}\ }\textbf {\bibinfo {volume} {105}},\ \bibinfo {pages}
  {044309} (\bibinfo {year} {2022})}\BibitemShut {NoStop}%
\bibitem [{\citenamefont {Lotze}\ \emph {et~al.}(2022)\citenamefont {Lotze},
  \citenamefont {Marszal}, \citenamefont {Schröder},\ and\ \citenamefont
  {Timme}}]{Lotze_2022}%
  \BibitemOpen
  \bibfield  {author} {\bibinfo {author} {\bibfnamefont {C.}~\bibnamefont
  {Lotze}}, \bibinfo {author} {\bibfnamefont {P.}~\bibnamefont {Marszal}},
  \bibinfo {author} {\bibfnamefont {M.}~\bibnamefont {Schröder}}, \ and\
  \bibinfo {author} {\bibfnamefont {M.}~\bibnamefont {Timme}},\ }\href
  {\doibase 10.1088/1367-2630/ac47c9} {\bibfield  {journal} {\bibinfo
  {journal} {New Journal of Physics}\ }\textbf {\bibinfo {volume} {24}},\
  \bibinfo {pages} {023034} (\bibinfo {year} {2022})}\BibitemShut {NoStop}%
\bibitem [{\citenamefont {{United States Census Bureau}}(2020)}]{NYC-census}%
  \BibitemOpen
  \bibfield  {author} {\bibinfo {author} {\bibnamefont {{United States Census
  Bureau}}},\ }\href@noop {} {\enquote {\bibinfo {title} {Quick{F}acts -- {N}ew
  {Y}ork {C}ity},}\ }\bibinfo {howpublished}
  {\url{https://www.census.gov/quickfacts/fact/table/newyorkcitynewyork/PST045221}}
  (\bibinfo {year} {2020}),\ \bibinfo {note} {accessed: 2022-10-04}\BibitemShut
  {NoStop}%
\bibitem [{\citenamefont {{NYC Taxi \& Limousine
  Commission}}(2022)}]{NYC-taxi}%
  \BibitemOpen
  \bibfield  {author} {\bibinfo {author} {\bibnamefont {{NYC Taxi \& Limousine
  Commission}}},\ }\href@noop {} {\enquote {\bibinfo {title} {{NYC Taxi and
  Limousine Commission Trip Record Data}},}\ }\bibinfo {howpublished}
  {\url{https://www1.nyc.gov/site/tlc/about/tlc-trip-record-data.page}}
  (\bibinfo {year} {2022}),\ \bibinfo {note} {accessed: 05-10-2022}\BibitemShut
  {NoStop}%
\bibitem [{\citenamefont {{New York City Department of
  Transportation}}(2018)}]{NYC-speed}%
  \BibitemOpen
  \bibfield  {author} {\bibinfo {author} {\bibnamefont {{New York City
  Department of Transportation}}},\ }\href@noop {} {\enquote {\bibinfo {title}
  {{New York City Mobility Report}},}\ }\bibinfo {howpublished}
  {\url{https://www1.nyc.gov/html/dot/html/about/mobilityreport.shtml}}
  (\bibinfo {year} {{2018}}),\ \bibinfo {note} {accessed:
  2022-10-05}\BibitemShut {NoStop}%
\bibitem [{\citenamefont {{Amt für Statistik
  Berlin-Brandenburg}}(2021)}]{BER-census}%
  \BibitemOpen
  \bibfield  {author} {\bibinfo {author} {\bibnamefont {{Amt für Statistik
  Berlin-Brandenburg}}},\ }\href@noop {} {\enquote {\bibinfo {title}
  {Bevölkerungsstand {J}ahresergebnisse},}\ }\bibinfo {howpublished}
  {https://www.statistik-berlin-brandenburg.de/a-i-3-j} (\bibinfo {year}
  {2021}),\ \bibinfo {note} {accessed: 2022-10-04}\BibitemShut {NoStop}%
\bibitem [{\citenamefont {Gerike}\ \emph {et~al.}(2018)\citenamefont {Gerike},
  \citenamefont {Hubrich}, \citenamefont {Ließke}, \citenamefont {Wittig},\
  and\ \citenamefont {Wittwer}}]{mobinstaedten}%
  \BibitemOpen
  \bibfield  {author} {\bibinfo {author} {\bibfnamefont {R.}~\bibnamefont
  {Gerike}}, \bibinfo {author} {\bibfnamefont {S.}~\bibnamefont {Hubrich}},
  \bibinfo {author} {\bibfnamefont {F.}~\bibnamefont {Ließke}}, \bibinfo
  {author} {\bibfnamefont {S.}~\bibnamefont {Wittig}}, \ and\ \bibinfo {author}
  {\bibfnamefont {R.}~\bibnamefont {Wittwer}},\ }\href
  {\url{https://tu-dresden.de/bu/verkehr/ivs/srv/ressourcen/dateien/SrV2018_Staedtevergleich.pdf?lang=en}}
  {\  (\bibinfo {year} {2018})}\BibitemShut {NoStop}%
\bibitem [{Wir(2018)}]{Wirtschaftsdienst2018}%
  \BibitemOpen
  \href {\doibase {10.1007/s10273-018-2339-y}} {\bibfield  {journal} {\bibinfo
  {journal} {Wirtschaftsdienst}\ }\textbf {\bibinfo {volume} {98}},\ \bibinfo
  {pages} {608} (\bibinfo {year} {2018})}\BibitemShut {NoStop}%
\bibitem [{\citenamefont {{Statistische Ämter des Bundes und der
  Länder}}(2020)}]{EMS-census}%
  \BibitemOpen
  \bibfield  {author} {\bibinfo {author} {\bibnamefont {{Statistische Ämter
  des Bundes und der Länder}}},\ }\href@noop {} {\enquote {\bibinfo {title}
  {{Regionalatlas Deutschland}},}\ }\bibinfo {howpublished}
  {https://regionalatlas.statistikportal.de/} (\bibinfo {year} {2020}),\
  \bibinfo {note} {accessed: 2022-10-04}\BibitemShut {NoStop}%
\bibitem [{\citenamefont {{Inst. f. Arbeitsmarkt- und
  Berufsforschung}}(2018)}]{IAB2018}%
  \BibitemOpen
  \bibfield  {author} {\bibinfo {author} {\bibnamefont {{Inst. f. Arbeitsmarkt-
  und Berufsforschung}}},\ }\href@noop {} {\bibfield  {journal} {\bibinfo
  {journal} {Kurzbericht}\ }\textbf {\bibinfo {volume} {10}} (\bibinfo {year}
  {2018})}\BibitemShut {NoStop}%
\bibitem [{\citenamefont {Debreu}(1959)}]{debreu1959}%
  \BibitemOpen
  \bibfield  {author} {\bibinfo {author} {\bibfnamefont {G.}~\bibnamefont
  {Debreu}},\ }\href@noop {} {\bibfield  {journal} {\bibinfo  {journal}
  {Proceedings of the National Academy of Sciences}\ }\textbf {\bibinfo
  {volume} {40}},\ \bibinfo {pages} {588} (\bibinfo {year} {1959})}\BibitemShut
  {NoStop}%
\bibitem [{\citenamefont {Greenwald}\ and\ \citenamefont
  {Stiglitz}(1959)}]{greenwald1986}%
  \BibitemOpen
  \bibfield  {author} {\bibinfo {author} {\bibfnamefont {B.}~\bibnamefont
  {Greenwald}}\ and\ \bibinfo {author} {\bibfnamefont {J.~E.}\ \bibnamefont
  {Stiglitz}},\ }\href@noop {} {\bibfield  {journal} {\bibinfo  {journal}
  {Quarterly Journal of Economics}\ }\textbf {\bibinfo {volume} {40}},\
  \bibinfo {pages} {229} (\bibinfo {year} {1959})}\BibitemShut {NoStop}%
\bibitem [{\citenamefont {Magill}\ and\ \citenamefont
  {Quinzii}(2002)}]{magill2002}%
  \BibitemOpen
  \bibfield  {author} {\bibinfo {author} {\bibfnamefont {M.}~\bibnamefont
  {Magill}}\ and\ \bibinfo {author} {\bibfnamefont {M.}~\bibnamefont
  {Quinzii}},\ }\href@noop {} {\emph {\bibinfo {title} {Theory of incomplete
  markets}}}\ (\bibinfo  {publisher} {MIT Press},\ \bibinfo {year}
  {2002})\BibitemShut {NoStop}%
\bibitem [{\citenamefont {Knörr}\ \emph {et~al.}(2016)\citenamefont {Knörr},
  \citenamefont {Heidt}, \citenamefont {Gores},\ and\ \citenamefont
  {Bergk}}]{TREMOD}%
  \BibitemOpen
  \bibfield  {author} {\bibinfo {author} {\bibfnamefont {W.}~\bibnamefont
  {Knörr}}, \bibinfo {author} {\bibfnamefont {C.}~\bibnamefont {Heidt}},
  \bibinfo {author} {\bibfnamefont {S.}~\bibnamefont {Gores}}, \ and\ \bibinfo
  {author} {\bibfnamefont {F.}~\bibnamefont {Bergk}},\ }\href@noop {} {\emph
  {\bibinfo {title} {Aktualisierung „{D}aten- und {R}echenmodell:
  Energieverbrauch und Schadstoffemissionen des motorisierten Verkehrs in
  Deutschland 1960-2035“ ({TREMOD}) für die Emissionsberichterstattung 2016
  (Berichtsperiode 1990-2014) Anhang}}},\ \bibinfo {type} {Tech. Rep.}\
  (\bibinfo {year} {2016})\BibitemShut {NoStop}%
\bibitem [{viz(2022)}]{viz2021}%
  \BibitemOpen
  \href@noop {} {\emph {\bibinfo {title} {{Verkehr in {Z}ahlen 2021/2022}}}},\
  \bibinfo {type} {Tech. Rep.}\ (\bibinfo  {institution} {{Bundesministerium
  für Verkehr und digitale Infrastruktur}},\ \bibinfo {year}
  {2022})\BibitemShut {NoStop}%
\bibitem [{spr(2022)}]{sprinter_energy}%
  \BibitemOpen
  \href@noop {} {\enquote {\bibinfo {title} {{Mercedez-Benz Configurator}},}\
  }\bibinfo {howpublished}
  {\url{https://voc.mercedes-benz.com/voc/de_de?_ga=2.230012379.695886780.1664812666-1473138929.1664499805}}
  (\bibinfo {year} {2022}),\ \bibinfo {note} {accessed:2022-10-03}\BibitemShut
  {NoStop}%
\bibitem [{\citenamefont {M\"uhle}(2022)}]{Muehle2022}%
  \BibitemOpen
  \bibfield  {author} {\bibinfo {author} {\bibfnamefont {S.}~\bibnamefont
  {M\"uhle}},\ }\href@noop {} {\emph {\bibinfo {title} {An analytical framework
  for modeling ride pooling efficiency and minimum fleet size}}}\ (\bibinfo
  {publisher} {under review},\ \bibinfo {year} {2022})\BibitemShut {NoStop}%
\bibitem [{\citenamefont {Salonen}\ and\ \citenamefont
  {Toivonen}(2013)}]{SALONEN2013143}%
  \BibitemOpen
  \bibfield  {author} {\bibinfo {author} {\bibfnamefont {M.}~\bibnamefont
  {Salonen}}\ and\ \bibinfo {author} {\bibfnamefont {T.}~\bibnamefont
  {Toivonen}},\ }\href {\doibase
  https://doi.org/10.1016/j.jtrangeo.2013.06.011} {\bibfield  {journal}
  {\bibinfo  {journal} {Journal of Transport Geography}\ }\textbf {\bibinfo
  {volume} {31}},\ \bibinfo {pages} {143} (\bibinfo {year} {2013})}\BibitemShut
  {NoStop}%
\bibitem [{\citenamefont {Liao}\ \emph {et~al.}(2020)\citenamefont {Liao},
  \citenamefont {Gil}, \citenamefont {Pereira}, \citenamefont {Yeh},\ and\
  \citenamefont {Verendel}}]{liao2020disparities}%
  \BibitemOpen
  \bibfield  {author} {\bibinfo {author} {\bibfnamefont {Y.}~\bibnamefont
  {Liao}}, \bibinfo {author} {\bibfnamefont {J.}~\bibnamefont {Gil}}, \bibinfo
  {author} {\bibfnamefont {R.~H.}\ \bibnamefont {Pereira}}, \bibinfo {author}
  {\bibfnamefont {S.}~\bibnamefont {Yeh}}, \ and\ \bibinfo {author}
  {\bibfnamefont {V.}~\bibnamefont {Verendel}},\ }\href@noop {} {\bibfield
  {journal} {\bibinfo  {journal} {Scientific Reports}\ }\textbf {\bibinfo
  {volume} {10}},\ \bibinfo {pages} {1} (\bibinfo {year} {2020})}\BibitemShut
  {NoStop}%
\bibitem [{\citenamefont {Fulman}\ and\ \citenamefont
  {Benenson}(2021)}]{fulman2021approximation}%
  \BibitemOpen
  \bibfield  {author} {\bibinfo {author} {\bibfnamefont {N.}~\bibnamefont
  {Fulman}}\ and\ \bibinfo {author} {\bibfnamefont {I.}~\bibnamefont
  {Benenson}},\ }\href@noop {} {\bibfield  {journal} {\bibinfo  {journal}
  {Transportation Science}\ }\textbf {\bibinfo {volume} {55}},\ \bibinfo
  {pages} {1046} (\bibinfo {year} {2021})}\BibitemShut {NoStop}%
\bibitem [{\citenamefont {Chaniotakis}\ and\ \citenamefont
  {Pel}(2015)}]{CHANIOTAKIS2015228}%
  \BibitemOpen
  \bibfield  {author} {\bibinfo {author} {\bibfnamefont {E.}~\bibnamefont
  {Chaniotakis}}\ and\ \bibinfo {author} {\bibfnamefont {A.~J.}\ \bibnamefont
  {Pel}},\ }\href {\doibase https://doi.org/10.1016/j.tra.2015.10.004}
  {\bibfield  {journal} {\bibinfo  {journal} {Transportation Research Part A:
  Policy and Practice}\ }\textbf {\bibinfo {volume} {82}},\ \bibinfo {pages}
  {228} (\bibinfo {year} {2015})}\BibitemShut {NoStop}%
\bibitem [{\citenamefont {Avermann}\ and\ \citenamefont
  {Schl\"uter}(2019)}]{Avermann2019}%
  \BibitemOpen
  \bibfield  {author} {\bibinfo {author} {\bibfnamefont {N.}~\bibnamefont
  {Avermann}}\ and\ \bibinfo {author} {\bibfnamefont {J.}~\bibnamefont
  {Schl\"uter}},\ }\href@noop {} {\bibfield  {journal} {\bibinfo  {journal}
  {Research in Transportation Business \& Management}\ }\textbf {\bibinfo
  {volume} {32}},\ \bibinfo {pages} {100420} (\bibinfo {year}
  {2019})}\BibitemShut {NoStop}%
\bibitem [{\citenamefont {Nyga}\ \emph {et~al.}(2020)\citenamefont {Nyga},
  \citenamefont {Minnich},\ and\ \citenamefont {Schl\"uter}}]{Nyga2020}%
  \BibitemOpen
  \bibfield  {author} {\bibinfo {author} {\bibfnamefont {A.}~\bibnamefont
  {Nyga}}, \bibinfo {author} {\bibfnamefont {A.}~\bibnamefont {Minnich}}, \
  and\ \bibinfo {author} {\bibfnamefont {J.}~\bibnamefont {Schl\"uter}},\
  }\href@noop {} {\bibfield  {journal} {\bibinfo  {journal} {Transportation
  Research A}\ }\textbf {\bibinfo {volume} {132}},\ \bibinfo {pages} {540}
  (\bibinfo {year} {2020})}\BibitemShut {NoStop}%
\bibitem [{\citenamefont {S\"orensen}\ \emph {et~al.}(2021)\citenamefont
  {S\"orensen}, \citenamefont {Bossert}, \citenamefont {Jokinen},\ and\
  \citenamefont {Schl\"uter}}]{Soerensen2021}%
  \BibitemOpen
  \bibfield  {author} {\bibinfo {author} {\bibfnamefont {L.}~\bibnamefont
  {S\"orensen}}, \bibinfo {author} {\bibfnamefont {A.}~\bibnamefont {Bossert}},
  \bibinfo {author} {\bibfnamefont {J.-P.}\ \bibnamefont {Jokinen}}, \ and\
  \bibinfo {author} {\bibfnamefont {J.}~\bibnamefont {Schl\"uter}},\
  }\href@noop {} {\bibfield  {journal} {\bibinfo  {journal} {Transport Policy}\
  }\textbf {\bibinfo {volume} {100}},\ \bibinfo {pages} {5} (\bibinfo {year}
  {2021})}\BibitemShut {NoStop}%
\end{thebibliography}%


\begin{thebibliography}{4}%
\makeatletter
\providecommand \@ifxundefined [1]{%
 \@ifx{#1\undefined}
}%
\providecommand \@ifnum [1]{%
 \ifnum #1\expandafter \@firstoftwo
 \else \expandafter \@secondoftwo
 \fi
}%
\providecommand \@ifx [1]{%
 \ifx #1\expandafter \@firstoftwo
 \else \expandafter \@secondoftwo
 \fi
}%
\providecommand \natexlab [1]{#1}%
\providecommand \enquote  [1]{``#1''}%
\providecommand \bibnamefont  [1]{#1}%
\providecommand \bibfnamefont [1]{#1}%
\providecommand \citenamefont [1]{#1}%
\providecommand \href@noop [0]{\@secondoftwo}%
\providecommand \href [0]{\begingroup \@sanitize@url \@href}%
\providecommand \@href[1]{\@@startlink{#1}\@@href}%
\providecommand \@@href[1]{\endgroup#1\@@endlink}%
\providecommand \@sanitize@url [0]{\catcode `\\12\catcode `\$12\catcode
  `\&12\catcode `\#12\catcode `\^12\catcode `\_12\catcode `\%12\relax}%
\providecommand \@@startlink[1]{}%
\providecommand \@@endlink[0]{}%
\providecommand \url  [0]{\begingroup\@sanitize@url \@url }%
\providecommand \@url [1]{\endgroup\@href {#1}{\urlprefix }}%
\providecommand \urlprefix  [0]{URL }%
\providecommand \Eprint [0]{\href }%
\providecommand \doibase [0]{http://dx.doi.org/}%
\providecommand \selectlanguage [0]{\@gobble}%
\providecommand \bibinfo  [0]{\@secondoftwo}%
\providecommand \bibfield  [0]{\@secondoftwo}%
\providecommand \translation [1]{[#1]}%
\providecommand \BibitemOpen [0]{}%
\providecommand \bibitemStop [0]{}%
\providecommand \bibitemNoStop [0]{.\EOS\space}%
\providecommand \EOS [0]{\spacefactor3000\relax}%
\providecommand \BibitemShut  [1]{\csname bibitem#1\endcsname}%
\let\auto@bib@innerbib\@empty
\bibitem [{\citenamefont {M\"uhle}(2022)}]{muehle2022}%
  \BibitemOpen
  \bibfield  {author} {\bibinfo {author} {\bibfnamefont {Steffen}\ \bibnamefont
  {M\"uhle}},\ }\href@noop {} {\emph {\bibinfo {title} {An analytical framework
  for modeling ride pooling efficiency and minimum fleet size}}}\ (\bibinfo
  {publisher} {under review},\ \bibinfo {year} {2022})\BibitemShut {NoStop}%
\bibitem [{\citenamefont {Horni}\ \emph {et~al.}(2016)\citenamefont {Horni},
  \citenamefont {Nagel},\ and\ \citenamefont {Axhausen}}]{matsim}%
  \BibitemOpen
  \bibinfo {editor} {\bibfnamefont {Andreas}\ \bibnamefont {Horni}}, \bibinfo
  {editor} {\bibfnamefont {Kai}\ \bibnamefont {Nagel}}, \ and\ \bibinfo
  {editor} {\bibfnamefont {Kay}\ \bibnamefont {Axhausen}},\ eds.,\ \href
  {\doibase 10.5334/baw} {\emph {\bibinfo {title} {Multi-Agent Transport
  Simulation MATSim}}}\ (\bibinfo  {publisher} {Ubiquity Press},\ \bibinfo
  {address} {London},\ \bibinfo {year} {2016})\ p.\ \bibinfo {pages}
  {618}\BibitemShut {NoStop}%
\bibitem [{tra(2022{\natexlab{a}})}]{train_NY}%
  \BibitemOpen
  \href@noop {} {\enquote {\bibinfo {title} {Wikipedia},}\ }\bibinfo
  {howpublished} {\url{https://en.wikipedia.org/wiki/New_York_City_Subway}}
  (\bibinfo {year} {2022}{\natexlab{a}}),\ \bibinfo {note} {accessed:
  26-10-2022}\BibitemShut {NoStop}%
\bibitem [{tra(2022{\natexlab{b}})}]{train_Berlin}%
  \BibitemOpen
  \href@noop {} {\enquote {\bibinfo {title} {Wikipedia},}\ }\bibinfo
  {howpublished} {\url{https://en.wikipedia.org/wiki/Berlin_U-Bahn}} (\bibinfo
  {year} {2022}{\natexlab{b}}),\ \bibinfo {note} {accessed:
  26-10-2022}\BibitemShut {NoStop}%
\end{thebibliography}%

\end{document}


\title{Supplemental Material for: Sustainable and convenient: bi-modal public transit systems outperforming the private car}
\author{Puneet Sharma}
\email{puneet.sharma@ds.mpg.de}
\author{Knut M. Heidemann}
 \email{knut.heidemann@ds.mpg.de}
\author{Helge Heuer}
\author{Steffen M\"uhle}
\author{Stephan Herminghaus}
\affiliation{Max-Planck Institute for Dynamics and Self-Organization (MPIDS), Am Fa\ss berg 17, 37077 G\"ottingen, Germany}


\maketitle


\section{Traffic volume}
To derive an expression for road traffic due to shuttles we connect the distance served by busses (left hand side) and the distance traveled by customers (right hand side), with average distance $\langle d \rangle$ traveled on road, for an area of reference $A$ during time $t_0=D/v_0$, with shuttle speed $v_0$
\begin{align}
\label{eq:continuity}
\underbrace{B A\, v_0 \occupancy\,\pdriving\, t_0}_{\textrm{distance\ served\ by\ busses}}  = \underbrace{\nu E A\, t_0 \,  \langle d \rangle \detour\,\paccept \, \,.}_{\textrm{distance\ traveled\ by\ customers}}
\end{align}
%
    Here $B$ is the shuttle density, $\occupancy$ is the average shuttle occupancy, $\nu E  A t_0$ is the number of requests in the reference area $A$ during reference time $t_0$ (with request frequency $\nu$, population density $E$), $\detour$ is the average relative detour of customers (no detour means $\detour = 1$), $p_\textrm{driving}$ is the probability of vehicles driving (i.e., not being idle), and $p_\textrm{accept}$ is the probability that a user request is accepted.
    In this study, we assume that all requests can be served ($\paccept=1$), i.e., no request has to be rejected because certain constraints (like a maximum waiting time) cannot be fulfilled (see~\cite{muehle2022} for a detailed analysis of $\paccept$).
    Substituting $\eta \equiv \occupancy/\detour$\footnote{See \cite{muehle2022} for details on why this identity holds.} in Eq.~\ref{eq:continuity}, we define the traffic volume, $\Gamma$, by the number of driving vehicles, i.e.:
    \begin{equation}\label{eq:gamma}
        \Gamma \equiv B A \cdot p_\textrm{driving} = \frac{\nu E \langle d \rangle A}{\eta v_0}\,.
    \end{equation}
    Note that for private cars, $\eta = 1$ and $\langle d \rangle \equiv D$, and for a bi-modal service, $\langle d \rangle = \Dbi $  and $\nu_\mathrm{shuttle} = (1+F)\nu$ (Eq.~10, 11 in the main manuscript). 
    Substituting these expressions in Eq.~\ref{eq:gamma}, we define the normalized traffic volume, $\tilde{\Gamma}$, as the ratio of traffic volume by shuttles in the bi-modal system, $\Gamma_\mathrm{bi}$, and the traffic volume by private cars, $\Gamma_\mathrm{MIV}$, i.e.,
    \begin{equation}\label{eq:traffic_bimodal}
        \tilde{\Gamma} \equiv \Gamma_\mathrm{bi} / \Gamma_\textrm{MIV} =\eta^{-1} (1+F) \,\tilde D_\mathrm{shuttle}\,.
    \end{equation}
    
    \section{DRRP pooling efficiency in bi-modal transport}
\begin{figure}
    \centering
    \includegraphics{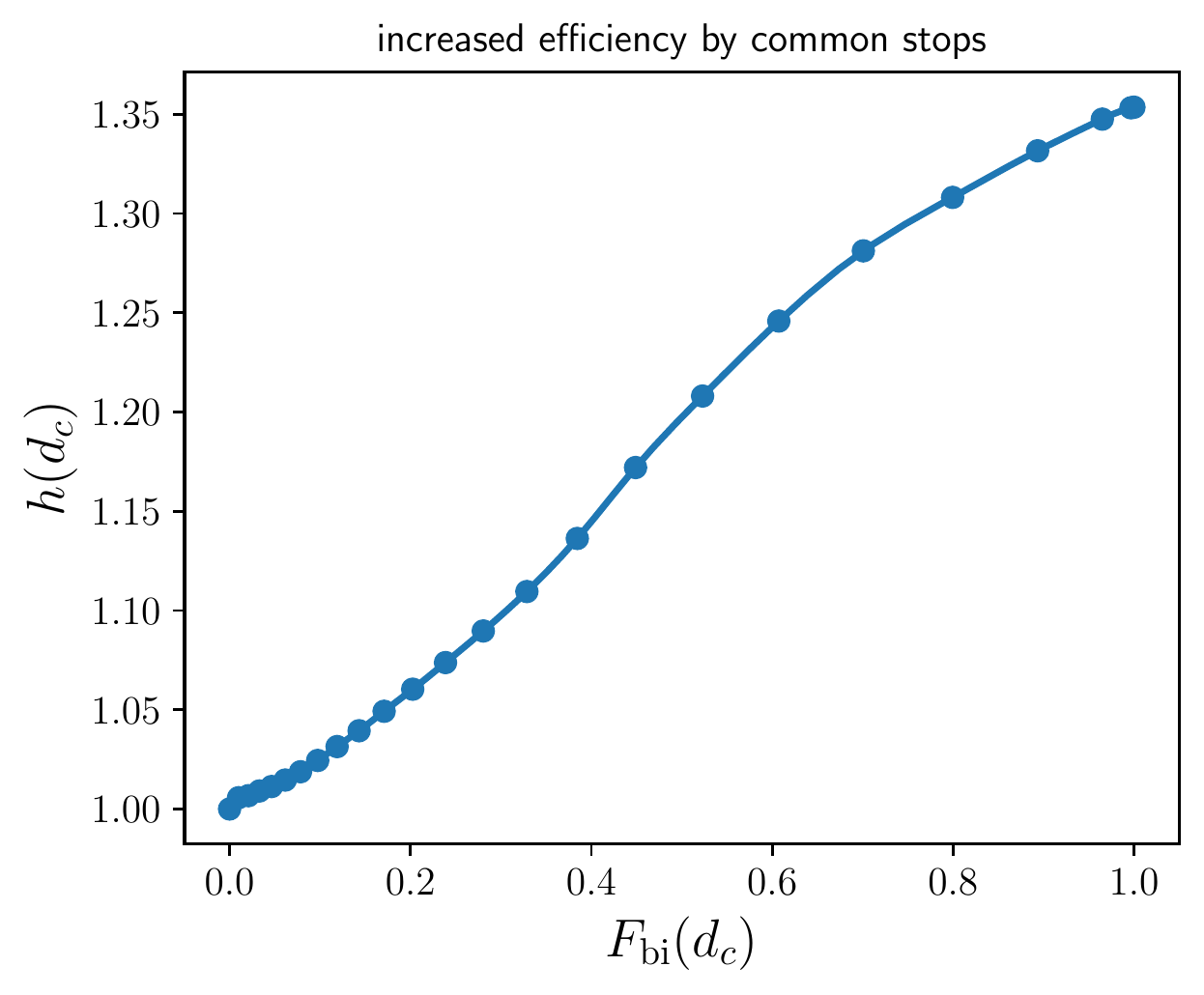}
    \caption{$h(F(\dcut))$ as a function of $F(\dcut)$. The 'common stop effect' is maximal for $F=1$, this is when all trips are served by bi-modal transport and all trips have either common origin or destination.}
    \label{fig:common_stop}
\end{figure}

    Simulations\cite{matsim} show that the DRRP pooling efficiency $\eta$ in a bi-modal system of transportation is larger than expected by the relation $\eta \propto \Lambda_\mathrm{shuttle}^\gamma$, with $\gamma\approx 0.12$, as derived in \cite{muehle2022}.
    This observation we attribute to what we call \emph{'common stop effect'}, meaning that pooling gets more efficient because bi-modal requests are spatially correlated due to shared pick-up and drop-off locations, i.e., the train stations.
    Here we present the empirical scaling function $h(F)$ accounting for this effect (see~Eq.~12 in the main manuscript), assuming that the 'common stop effect' is only governed by the fraction $F$ of people assigned to bi-modal transportation.
    We plot the empirical data for $h(F)$ as a function of $F$ in Fig.~\ref{fig:common_stop}. As expected the 'common stop effect' is maximal for $F = 1$.
    
    \section{Waiting time ($\tau_{w}$) and detour ($\delta$)}
\begin{figure}
    \centering
    \includegraphics[width=\linewidth]{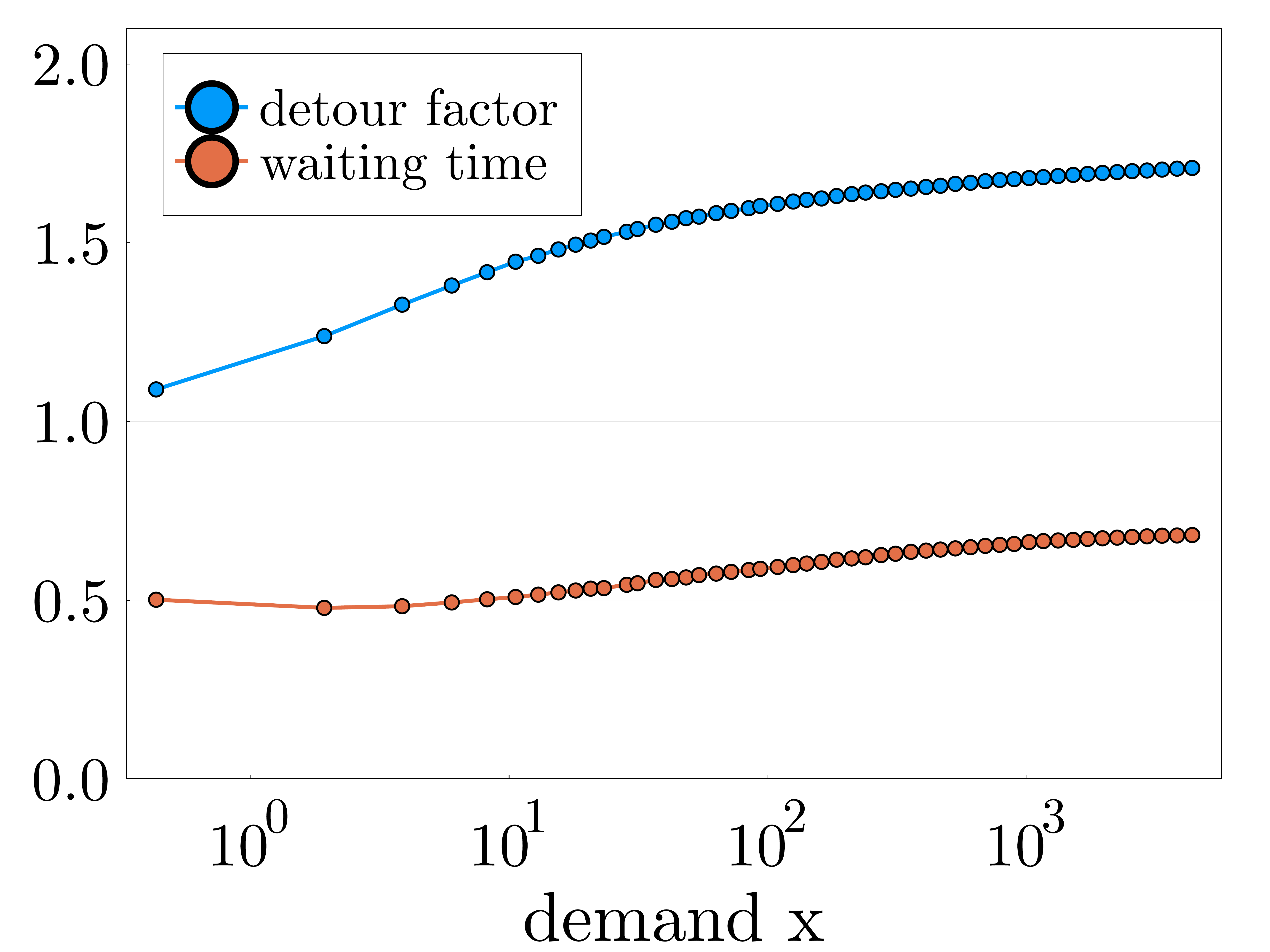}
    \caption{Detour factor $(\delta)$ and mean waiting time $(\tau_w)$ for various demands.}
    \label{fig:delta_tw}
\end{figure}

We use simulation results to determine the average waiting time for shuttles, $\tau_w$, and the average detour, $\delta$.
In Fig.~\ref{fig:delta_tw}, mean detour and mean waiting time are plotted as a function of user demand.
Users are assumed to have a maximum accepted waiting time of $t_{0}$.
We observe that mean waiting time $\tau_w$ grows with demand and saturates around $0.65$.
Similarly the mean detour $\delta$ grows with demand and saturates around $1.65$.


\section{Decision on transport service type}
We compare the time it takes to serve a new user request $(\mathcal{P},\mathcal{D})$ by bi-modal and uni-modal transportation. In order to do so we sample $N=10^5$ pick-up $(\mathcal{P})$ and drop-off $(\mathcal{D})$ pairs. The trains operate at $\tilde{\mu}=2$.

The time it takes to serve a randomly sampled user request $(\mathcal{P},\mathcal{D})$ by bi-modal transportation $(t_\textrm{bi})$ comprises of driving time in two shuttles to the nearest train station, driving time in train between the two train stations, waiting time due to shuttles and waiting time at the train station. If $\mathcal{S}_\mathcal{P}$ and $\mathcal{S}_\mathcal{D}$ represent the location of the train station next to $\mathcal{P}$ and $\mathcal{D}$, respectively, then
\begin{equation}
    t_\textrm{bi} = \underbrace{2t^\textrm{shuttle}_w + \delta\frac{\overline{\mathcal{P}\mathcal{S}_\mathcal{P}}}{v_0} + \delta\frac{\overline{\mathcal{D}\mathcal{S}_\mathcal{D}}}{v_0}}_\textrm{two shuttle trips} + \underbrace{t^\textrm{train}_{w}+\frac{\overline{\mathcal{S}_\mathcal{P}\mathcal{S}_\mathcal{D}}}{v_\textrm{train}}}_\textrm{train}\,,
\end{equation}
where $t^\textrm{shuttle}_w,t^\textrm{train}_w$ are the waiting times incurred due to shuttles and trains and are assumed to take the values $t_0/2$ and $1/2\mu$, respectively.

The time taken to serve the same request by uni-modal transportation (shuttles only) is:
\begin{equation}
    t_\textrm{uni} = t^\textrm{shuttle}_w + \delta\frac{\overline{\mathcal{P}\mathcal{D}}}{v_0}\,.
\end{equation}
The ratio, $t_\textrm{bi}/t_\textrm{uni}$ is plotted in Fig.2a in the main text. 

In order to obtain Fig.2b, we plot the ratio of increment in energy usage for serving a new user request. 

The increment in energy usage when serving a new request by bi-modal transportation comprises of the increment in energy usage by two shuttle trips,
\begin{equation}
(\Delta\mathcal{E})_\textrm{bi} = \eta^{-1}\cdot e_\textrm{shuttle}\cdot(\overline{\mathcal{P}\mathcal{S}_\mathcal{P}} + \overline{\mathcal{D}\mathcal{S}_\mathcal{D}})\,.
\end{equation}

The increment in energy usage when served by uni-modal transportation comprises of only a single shuttle trip,
\begin{equation}
    (\Delta\mathcal{E})_{\textrm{uni}} =  \eta^{-1}\cdot e_\textrm{shuttle}\cdot\overline{\mathcal{P}\mathcal{D}}.
\end{equation}
The ratio $(\Delta\mathcal{E})_\textrm{bi}/(\Delta\mathcal{E})_{\textrm{uni}}$, i.e., $\frac{\overline{\mathcal{P}\mathcal{S}_\mathcal{P}} + \overline{\mathcal{D}\mathcal{S}_\mathcal{D}}}{\overline{\mathcal{P}\mathcal{D}}}$ is plotted in Fig.~2b in the main text.
This ratio, if greater than 1 (less than 1), indicates whether the two trips to/from the train stations are longer (shorter) than a direct trip.

Note that only shuttles contribute to $\Delta\mathcal{E}$ because a single user request is assumed to have no effect on train operations.

\section{Dependence of train speed on inter-station distance.}
Trains are assumed to have a maximum operating speed of $v_m$, acceleration and deceleration time $t_a$ and a stop time of $t_s$ at every station.
The effective average train speed is therefore
\begin{equation}\label{eq:train_speed}
    v_\textrm{train} = 
    \begin{cases}
    
    \frac{\ell}{\frac{\ell}{v_m} + t_a + t_s}, & \textrm{if } \ell \geq v_m\cdot t_a \\
    \frac{\ell}{2\sqrt{\frac{\ell t_a}{v_m}}+t_s}, &\textrm{otherwise}
    \end{cases}\,.
\end{equation}
For New York, we use $v_m = 89\,\si{km/h}$ and $v_\textrm{train} = 28\,\si{km/h}$ \cite{train_NY} at $\tilde{\ell}=0.28$ (see Tab.~1 in the main text).
We use Eq.~\ref{eq:train_speed} to determine $t_a$ and $t_s$ by assuming that $t_a = t_s$.
Similarly, for Berlin we use $v_m = 72\,\si{km/h}$ and $v_\textrm{train}=30.7\,\si{km/h}$ \cite{train_Berlin} at $\tilde{\ell}=0.32$.
For our analysis, we use New York train speed as a proxy for $\Lambda = 10^4$ and Berlin train speed for $\Lambda = \{10^3, 10^2\}$.

\bibliography{Bimodal_si}